%% file: Wilson-coefficients.tex
\documentclass[a4paper,11pt]{article}
\pdfoutput=1 

\usepackage{jheppub} 

\usepackage[T1]{fontenc} 
\usepackage{graphics}
\usepackage{amssymb}
\usepackage{amsmath}
\usepackage{bbold}
\usepackage{hyperref}
\usepackage{slashed}
\usepackage[normalem]{ulem}
\usepackage{booktabs}
\usepackage{orcidlink}  

\usepackage{tikz}
\usetikzlibrary{positioning,arrows}
\usepgflibrary{arrows.meta}
\usetikzlibrary{decorations.pathmorphing}
\usetikzlibrary{decorations.markings}
\usetikzlibrary{math}
\usetikzlibrary{calc}

\tikzset{
gluon/.style={thick,decorate, draw=black, decoration={coil,aspect=0.7, post length = 0pt, pre length = 0pt, segment length=3pt,amplitude=2pt}},
quark1/.style={thick, postaction={decorate}, decoration={markings,mark=at position .5 with {\arrow{Stealth[scale=1]}}}},
quark2/.style={thick, postaction={decorate}, decoration={markings,mark=at position .75 with {\arrow{Stealth[scale=1]}}}},
cross/.style={path picture={
  \draw[black,thick]
(path picture bounding box.south east) -- (path picture bounding box.north west) (path picture bounding box.south west) -- (path picture bounding box.north east);
}}
}

\allowdisplaybreaks

\newcommand{\ord}[1]{{\mathcal O}(#1)}

\newcommand{\msb}{{\overline{\mathrm{MS}}}}
\newcommand{\eps}{\epsilon}
\newcommand{\nn}{\nonumber}

\newcommand{\as}{\alpha_s}

\DeclareRobustCommand{\eq}[1]{eq.~\eqref{eq:#1}}
\DeclareRobustCommand{\eqs}[2]{eqs.~\eqref{eq:#1} and \eqref{eq:#2}}
\DeclareRobustCommand{\eqsm}[2]{eqs.~\eqref{eq:#1}\,--\,\eqref{eq:#2}}

\DeclareRobustCommand{\fig}[1]{figure~\ref{fig:#1}}

\DeclareRobustCommand{\sec}[1]{section~\ref{sec:#1}}

\DeclareRobustCommand{\subsec}[1]{section~\ref{subsec:#1}}

\DeclareRobustCommand{\app}[1]{appendix~\ref{app:#1}}

\DeclareRobustCommand{\rcite}[1]{ref.\,\cite{#1}}
\DeclareRobustCommand{\rcites}[1]{refs.\,\cite{#1}}
\DeclareRobustCommand{\tab}[1]{table~\ref{tab:#1}}

\newcommand{\ri}{\mathrm{i}}

\newcommand{\rd}{\mathrm{d}}
\newcommand{\rnd}{\mathrm{nd}}
\newcommand{\re}{\mathrm{e}}
\newcommand{\rT}{\mathrm{T}}
\newcommand{\rs}{\mathrm{s}}
\newcommand{\rns}{\mathrm{ns}}
\newcommand{\bare}{\mathrm{bare}}
\newcommand{\sub}{\mathrm{sub}}

\newcommand{\tr}{\mathrm{tr}}
\newcommand{\op}{\mathcal{O}}
\newcommand{\C}{\mathcal{C}}
\newcommand{\Q}{\mathrm{Q}}
\newcommand{\lOP}[1]{\lim_{#1}}
\newcommand{\CO}{\mathrm{[\C \op]}}

\newcommand{\dFFF}{d^{(3)}_{FF}}
\newcommand{\Rra}{\rho_1}
\newcommand{\Rrb}{\rho_2}
\newcommand{\Rrc}{\Q}
\newcommand{\Mm}{M}
\newcommand{\Mmp}[1]{M^{#1}}

\newcommand{\aop}{\frac{\alpha_s}{\pi}}
\newcommand{\aopn}[1]{\Bigl(\frac{\alpha_s}{\pi}\Bigr)^{\! #1}}

\DeclareMathAlphabet{\mathbbold}{U}{bbold}{m}{n}
\newcommand{\bbid}{\mathbbold{1}}

\title{
Three-Loop OPE Wilson Coefficients of Dimension-Four Operators for (Axial-)Vector and (Pseudo-)Scalar Current Correlators
}

\preprint{
\begin{flushright}
FR-PHENO-2024-007\\
UWThPh-2024-15
\end{flushright}
}

\author[a]{Robin Br\"user,}
\author[b]{Andr\'e H. Hoang\orcidlink{0000-0002-8424-9334} and} 
\author[a]{Maximilian Stahlhofen\orcidlink{0000-0002-2613-9014}}

\affiliation[a]{Albert-Ludwigs-Universit\"at Freiburg, Physikalisches Institut, D-79104 Freiburg, Germany}
\affiliation[b]{University of Vienna, Faculty of Physics, Boltzmanngasse 5, A-1090 Wien, Austria}

\emailAdd{robin.brueser@physik.uni-freiburg.de}
\emailAdd{andre.hoang@univie.ac.at}
\emailAdd{maximilian.stahlhofen@physik.uni-freiburg.de}

\abstract{
We calculate the three-loop Wilson coefficients of all physically relevant dimension-four operators, i.e.\ $G_{\mu\nu}^a G^{a,\mu\nu}$, $m_i\bar q_j q_j$ and $m_i m_j m_k^2$,
in the short-distance expansion of the time-ordered product of a pair of gauge-singlet
vector, axial-vector, scalar and pseudo-scalar currents.
The results are given for a general non-Abelian  gauge theory with arbitrary (compact semi-simple) gauge group and $n_f$ light fermion flavors (quarks)  in a common arbitrary representation of the gauge group, which includes QCD as a special case.
In particular, we allow for arbitrary flavor contents of each of the currents.
For the axial-vector current the included contributions from so-called singlet diagrams are consistent with the one-loop axial anomaly.
}

\begin{document}
\maketitle
\flushbottom

\section{Introduction}
\label{sec:intro}

The operator product expansion at short-distances (OPE)~\cite{Wilson:1969zs} represents one of the most fundamental concepts used in quantum chromodynamics (QCD).
It is employed in a multitude of phenomenological high-precision predictions where perturbative as well as non-perturbative corrections are mandatory.
One of the classic applications of the OPE,
which was pioneered a long time ago by Shifman, Vainshtein, and Zakharov~\cite{Shifman:1978bx,Shifman:1978by},
concerns the time-ordered product of two quark currents for the description of inclusive hadron production rates at large energies.
For currents with mass dimension three the corresponding OPE has the schematic form
\begin{align}
\label{eq:opedef}
\!\int\!\! \rd^4 x \; \re^{\ri q  \cdot x} \;
    \rT \{j_{1}(x) \, j_{2}(0) \} =
    \sum_{k=0,1,2,\ldots} \! \frac{1}{(-q^2)^{k-1}} \sum_n C^{(2k)}_n(q^2)\, \op^{(2k)}_n(0) \,,
\end{align}
where the $\op^{(m)}_n$ denote local gauge-invariant operators of gluon and light (up, down and strange) quark fields
with mass dimension $m$. The operators are multiplied by their Wilson coefficients $C^{(m)}_n$, which are defined to be dimensionless functions in \eq{opedef}.
The OPE terms are ordered hierarchically in the operator dimension multiplied by corresponding inverse powers of the squared hadronic invariant mass $q^2$.
For vacuum correlators the OPE yields a series of vacuum operator matrix elements $\langle \op^{(m)}_n \rangle$, generically called ``condensates''.
Important state-of-the-art high-precision applications of this OPE involve the description of inclusive semileptonic $\tau$ decays~\cite{Pich:1999hc} relevant for strong coupling~\cite{Pich:2016bdg,Boito:2020xli,Ayala:2022cxo} and strange quark mass determinations~\cite{Ananthanarayan:2016kll,Bodenstein:2013paa}, as well as QCD sum rules to determine charm and bottom quark masses from the hadronic $R$-ratio in electron-positron annihilation in the continuum~\cite{Bodenstein:2011ma,Dehnadi:2015fra,Erler:2016atg,Chetyrkin:2017lif}
or the quarkonium region~\cite{Hoang:2000fm,Hoang:2012us,Beneke:2014pta}.

At dimension four there are three types of vacuum matrix elements that are physically relevant, the so-called gluon condensate
$\langle G_{\mu \nu}^a G^{a,\mu \nu}\rangle$, the quark condensates of the form $m_i \langle \bar{q}_j q_j \rangle$ and the pure quark mass
terms of the form $m_i m_j m_k^2$, where the indices $i,j,k$ stand for (not necessarily different) light quark flavors. Among these the quark condensates, which are related to chiral symmetry breaking~\cite{Gasser:1983yg,Jamin:2002ev}, are known with the highest precision and can also be determined from lattice QCD~\cite{FlavourLatticeAveragingGroupFLAG:2021npn}. The pure quark mass terms are known quite well too as they appear in many lattice calculations and perturbative sum rule analyses, see \rcite{ParticleDataGroup:2022pth}. The value of the gluon condensate is only roughly known. This is because the gluon condensate is rather difficult to control in lattice QCD, and also because there is an associated infrared (IR) renormalon in the dimension-zero perturbation series for the current correlator, which makes precise phenomenological determinations somewhat unreliable. Thus, the first determination from \rcites{Shifman:1978bx,Shifman:1978by} is frequently still considered state-of-the-art, see e.g.\ \rcite{Gubler:2018ctz} for a recent review.
As far as the Wilson coefficients of these dimension four operators are concerned, for many phenomenological applications their two-loop results, already obtained many years ago in \rcites{Chetyrkin:1985kn,Surguladze:1990sp}, are adequate.
For example, precise strong coupling determinations from inclusive semileptonic $\tau$ decays~\cite{Pich:2016bdg,Boito:2020xli,Ayala:2022cxo} can be achieved by considering only spectral function moments of the inclusive hadronic invariant mass spectrum where the non-perturbative corrections arising from the dimension-four operators are strongly suppressed.

In the recent articles~\cite{Benitez-Rathgeb:2022yqb,Benitez-Rathgeb:2022hfj}, concerning the computation of spectral function moments for inclusive semileptonic $\tau$ decays,
it has been pointed out that the large discrepancy between the common fixed-order method and the so-called contour-improved approach~\cite{Pivovarov:1991rh,LeDiberder:1992jjr} (frequently referred to as FOPT and CIPT, respectively) for the computation of the dominant dimension-zero Wilson coefficient can be removed by adopting a certain scheme for the gluon condensate.
In that scheme the associated IR renormalon is cancelled explicitly from the perturbation series of the dimension-zero Wilson coefficient and the gluon condensate becomes renormalon-free.
An alternative scheme was discussed in \rcite{Beneke:2023wkq}.
The effect is large in CIPT even for spectral function moments for which the corrections
from dimension-four OPE operators are supposed to be suppressed. This can be explained from the fact that the perturbative analytic structure inherent to the CIPT approach is not compatible with the canonical formulation of the OPE displayed in \eq{opedef}~\cite{Hoang:2020mkw,Hoang:2021nlz}.
Using such a renormalon-free scheme for the gluon condensate, its Wilson coefficient directly enters the dominant perturbation series proportional to the identity operator at leading order in the OPE, which is known up to $\ord{\alpha_s^4}$~\cite{Baikov:2008jh}.
Given that the Gluon condensate Wilson coefficient is so far only known at two loops, this reintroduces an $\ord{\alpha_s^3}$ uncertainty for the perturbative predictions of the spectral function moments in the renormalon-free gluon condensate scheme. This motivates to account for the higher-order corrections to the gluon condensate Wilson coefficient. Interestingly,
the three-loop Wilson coefficients of the dimension-four operators have been determined in the PhD thesis of R.~Harlander~\cite{HarlanderDiss} more than 25 years ago for flavor off-diagonal (non-singlet) currents. However, with the exception of the Wilson coefficients for the $m_i m_j m_k^2$ operators given in \rcite{Harlander:2020duo} the results have not yet been published in a scientific journal nor, to the best of our knowledge, used in phenomenological analyses. The main purpose of the present work is to validate these results in an independent calculation and to extend them to arbitrary flavor configurations, i.e.\ to provide new three-loop (and partly new lower-loop) results also for all flavor diagonal currents.

In this article we compute the three-loop Wilson coefficients of the above mentioned physically relevant dimension-four operators in the OPE of current correlators involving a pair of (axial-)vector and (pseudo-)scalar quark currents.
We verify the corresponding results for two flavor non-diagonal (non-singlet) currents  given in \rcites{HarlanderDiss,Harlander:2020duo} in the context of QCD (with $N_c=3$).
In addition to these previously determined results, we also calculate the Wilson coefficients for the cases where the currents are flavor singlets or of the non-singlet diagonal type for which we partially even obtain novel results at the two-loop level.
Interestingly, the quark condensates and the mass correction terms receive non-zero Wilson coefficients also in the cases of mixed flavor singlet and diagonal non-singlet or different diagonal non-singlet currents since the quark condensate operators and the quark masses act as additional flavor sources or sinks.
The flavor singlet current results necessarily involve additional (partly finite) renormalization factors for axial-vector and pseudo-scalar currents to reconcile anomaly and symmetry relations with the chosen scheme to treat $\gamma_5$ in dimensional regularization.
The finiteness and consistency of these results provide some of the non-trivial cross checks of our computations.
We provide the results in a very convenient and
compact form applicable for any current flavor content, which to the best of our knowledge has not been employed before in this context.
It allows for a straightforward extraction of the Wilson coefficient for any current flavor configuration of interest and the easy identification of the contributions from different diagram classes.
We hope that the new results we present in this way will be useful for further applications.
For the interested reader we also provide some technical details on how we set up our calculations.

The article is organized as follows: In \sec{ope} we define our notations and the operator basis.
Section~\ref{sec:calculation} provides details on the projections and computations of Feynman diagrams we employed to determine the Wilson coefficients for the physical dimension-four operators.
In \sec{renormalization} we explain the renormalization of the currents and the operators, which are subject to mixing.
Our results are discussed in general form in \sec{results}, and for some relevant current configurations explicitly in \sec{discussion}, where we also compare to the literature.
In \sec{conclusions} we conclude and give a brief outlook.
Our article includes two appendices showing the explicit expressions for all our results and, for the convenience of the reader, also the expressions of all required renormalization constants.
We also provide an ancillary file with our results in \texttt{Mathematica} readable format.

\section{Operator product expansion}
\label{sec:ope}

We consider the mass dimension-four (dim-4)
contributions to the OPE of the correlator
\begin{align}
    \label{eq:correlatordef}
    T_{(\mu\nu)}^J \equiv
    \ri  \!\int\!\! \rd^d x \; \re^{\ri q  \cdot x} \;
    \rT \Big\{j^J_{1(,\mu)}(x) \, j^J_{2(,\nu)}(0)
    \Big\}
\end{align}
of vector ($V$), axial-vector ($A$), scalar ($S$), and pseudo-scalar ($P$)  gauge-singlet  light quark currents defined by
\begin{align}
\label{eq:currentsdef}
  j_{i,\mu}^V = \bar{q} \, \gamma_\mu \, \rho_i  \, q \,, \quad
  j_{i,\mu}^A = \bar{q} \, \gamma_\mu \gamma_5 \, \rho_i \, q \,, \quad
  j_{i}^S = \bar{q} \, \rho_i \, q \,, \quad
  j_{i}^P = \ri\, \bar{q} \, \gamma_5 \, \rho_i \, q \,.
\end{align}
The quark field operators $q = (u,d,s,\ldots)^\mathrm{T}$ represent vectors in the $n_f$-dimensional flavor space.
The in general complex-valued ($n_f \times n_f$) matrices $\rho_i$ determine the flavor content of the currents
and essentially represent density matrices in flavor space.
For a Hermitian flavor matrix the corresponding
current is Hermitian as well.
Via a
Fierz-type identity each of the two  matrices $\rho_i$ can be decomposed into a (flavor) SU($n_f$) singlet part, proportional
to the unit matrix, and a non-singlet part, related to the SU($n_f$) generators $t^a$ ($a=1,\ldots, n_f^2-1$) in the
fundamental representation:
\begin{align}
\label{eq:fierz}
  \rho_i = \frac{1}{n_f} \tr[\rho_i] \,\bbid_{n_f \times n_f} + \frac{1}{T_F^f} \tr[\rho_i \, t^a]\, t^a\,.
\end{align}
While the singlet term is flavor diagonal entirely, the non-singlet term has both flavor non-diagonal and diagonal contributions. Examples for $n_f=3$ are the generators $t^{1,2,4,5,6,7}$ and $t^{3,8}$, respectively, using the standard convention where they are proportional to the Gell-Mann matrices.
The constant $T_F^f \equiv \tr[t^a t^a] /(n_f^2-1)$
fixes the norm of the flavor generators,
which can in principle be chosen independent of the corresponding norm for the gauge group generators. Here and in the rest of the article repeated adjoint indices are always understood to be summed.
We emphasize that the results we present below are valid for any choice of the flavor matrices
$\rho_1$ and $\rho_2$, i.e.\ we do not impose the condition  $\rho_2^\dagger = \rho_1$.
This means we also allow for mixed singlet--non-singlet currents and different non-singlet currents.
These also arise at (mass) dimension four since in flavor space the quark fields and masses of the operators  themselves act as flavor space density
matrices representing sources and sinks of flavor.
We also note that our results are valid for arbitrary flavor number $n_f$.
For concreteness we will assume the gauge (``color'') group to be SU($N_c$) and all fermions to transform in the fundamental color representation ($F$) throughout this paper.
This allows to easily specialize to QCD by setting $N_c=3$.
We stress, however, that it is straightforward to generalize all analytic expressions given below to an arbitrary (compact semi-simple) gauge group and an arbitrary irreducible fermion representation $R$ with dimension $N_R$ by replacing
$T_F \to T_R$, $N_c \to N_R$, $C_F \to C_R$, and $d^{(3)}_{FF} \to d^{(3)}_{RR}$.
In any case, we frequently adapt QCD terminology, i.e.\ call all fermions ``quarks'', the gauge boson ``gluon'', and the charge of the gauge group ``color''.

We parametrize the dim-4 terms in the OPE of the current correlators ($J=V,A,S,P$) arising in the Euclidean limit
$-q^2 \to \infty$ in the form
\begin{align}
  \label{eq:OPE}
  \left[T^{J}_{(\mu\nu)}\right]^{\rm dim-4} =
   \frac{1}{-q^2} \sum_n
   \C_{n(,\mu\nu)}^{J}(q^2,\mu)\,  \op_n(\mu) \,,
\end{align}
which implies that the Wilson coefficients are dimensionless.
For the vector and axial-vector current correlators ($J=V,A$) the Wilson coefficients can be further decomposed in a transversal ($T$) and a longitudinal ($L$) part. We define
\begin{align}
 \label{eq:WC-TL}
 \C_{n,\mu\nu}^{J}(q^2,\mu) = \frac{q_\mu q_\nu-q^2 g_{\mu\nu}}{-q^2}\,  \C_n^{J,T}(q^2,\mu) +
  \frac{q_\mu q_\nu}{-q^2}  \, \C_n^{J,L}(q^2,\mu) \,.
\end{align}
Throughout this paper, especially in the discussions of \sec{renormalization}, we will frequently suppress the Lorentz indices $\mu$, $\nu$, the labels $T$, $L$ of the vector and axial-vector Wilson coefficients as well as the $q^2$ and $\mu$ (renormalization scale) arguments of Wilson coefficients and operators for brevity of notation.
Here, the index $n$ generically labels the linearly independent operators $\op_n$ in some given basis.
A concrete choice for the dim-4 composite operator basis
is~\cite{Kluberg-Stern:1974iel,Collins:1976yq,Spiridonov:1984br,Chetyrkin:1985kn,HarlanderDiss}
\begin{align}
  \op_1 &= G_{\mu \nu}^a G^{a,\mu \nu}\,, \qquad
  \op_2^{ij} = m_i \bar{q}_j q_j \,, \qquad
  \op_3^i = \bar{q}_i  \biggl( \ri \overset{\leftrightarrow}{\slashed{D}}/2 - m_i \biggr) q_i \,, \nn\\
  \op_4 &= A_\nu^a \Bigl( D^{ab}_\mu G^{b,\mu\nu} + g \, \bar{q}\, T^a
  \gamma^\nu q \Bigr) - \partial_\mu \bar{c}^{\, a} \partial^\mu c^a\,,  \qquad
  \op_5 = \bigl( D^{ab}_\mu \partial^\mu \bar{c}^{\,b} \bigr) c^a\,, \nn\\
  \op_6^{ijk} &= m_i m_j m_k^2 \,\bbid \big|_{i\neq j} \,,  \qquad
  \op_6^{ij} = m_i^2 m_j^2 \,\bbid
  \label{eq:Ops}
\end{align}
where the covariant derivatives for the adjoint and
fermion (in QCD: fundamental) representations of the gauge group
are defined by\footnote{Notice our notation $T^a$ for the
color generators as opposed to $t^a$ for the SU($n_f$) generators.}
\begin{align}
  D_\mu^{ab} = \delta^{ab}\partial_\mu -  g f^{abc}A_\mu^c\,,
  \qquad
  \overset{\leftrightarrow}{D}_\mu =   \overset{\leftrightarrow}{\partial}_\mu - \ri g A^a_\mu T^a\,,
  \qquad
  \bar{q}  \overset{\leftrightarrow}{\partial}_\mu  q =
  \bar{q} (\partial_\mu  q) - ( \partial_\mu \bar{q}) q \,.
\end{align}
This operator basis is complete and contains apart from the physical operators $\op_1$, $\op_2^{ij}$,
$\op_6^{ijk}$, and $ \op_6^{ij}$
also operators that either vanish by the classical equations of motions or are gauge-dependent ($\op_3^i$, $\op_4$, $\op_5$). The latter do not contribute in the OPE for physical quantities. However, all operators are in general required to carry out consistent (off-shell) matching calculations. The aim of this paper is to compute the Wilson coefficients
$\C_1^{J,\mu\nu}$, $ \C_2^{J,ij}$, $\C_6^{J,ijk}$, and $\C_6^{J,ij}$ of the physical operators in \eq{Ops} for $J=V,A,S,P$.

We note that dim-4 operators of the form  $m_i m_j m_k m_l \bbid$ with mutually differing flavor indices $i,j,k,l$ are absent in the OPE, also for $n_f \ge 4$. From the perspective of the OPE
current correlator matching calculation, see \subsec{projection},
this can be understood from the fact that the two external currents $j_{1,2}^J$ can yield linear dependence on at most two different quark masses
upon expansion in small $m_i$.
Additional quark mass dependence can only arise from extra closed (single-flavor) quark loops.
The Dirac traces associated with these loops are, however, only non-vanishing for an even number of Dirac matrices, which implies an even power in the quark mass expansion.
From the perspective of the renormalization of the operators and the consequent mixing between them due to UV divergences, see \sec{renormalization},
the absence of $m_i m_j m_k m_l \bbid$ operators
can be seen from the fact that, again, linear dependence on at most two different quark masses can arise from the operators themselves.
All additional quark mass dependence can only arise from extra closed quark loops in diagrams beyond leading order, which leads to the same conclusion.

\section{Calculation of the Wilson coefficients}
\label{sec:calculation}

\subsection{Projecting onto the Wilson coefficients}
\label{subsec:projection}

\begingroup
\begin{table}[t]
\centering
 \begin{tabular}{ccl}
 \toprule
  Operator       & Vertex &\multicolumn{1}{c}{Feynman rule} \\\midrule
  $\op_1$        &
  \begin{tikzpicture}[scale=1, baseline=(current bounding box.center)]
  \draw[gluon] (-1,0) -- (-0.1,0) node[pos=0.2, above=1pt] {\footnotesize $a,\alpha$};
  \fill [black] (-0.1,-0.1) rectangle (0.1,0.1);
  \draw[gluon] (0.1,0) -- (1,0) node[pos=0.8, above=1pt] {\footnotesize $b,\beta$};
  \draw[->] (-0.9,-0.25) -- (-0.4,-0.25) node[pos=0.5, below=0.5pt] {\footnotesize $p_1$};
  \draw[->] (0.4,-0.25) -- (0.9,-0.25) node[pos=0.5, below=0.5pt] {\footnotesize $p_2$};
  \end{tikzpicture}
  &
  $\quad 4 \delta^{ab} \big [ p_1 \!\cdot\! p_2 \,g^{\alpha\beta} - p_1^\alpha p_2^\beta \big] $
  \\[4ex]
  $\op_2^{ij}$   &
   \begin{tikzpicture}[scale=1]
  \draw[quark1] (-1,0) -- (0,0) node[pos=0.15, above=1pt] {\footnotesize $f$};
  \fill [black] (-0.1,-0.1) rectangle (0.1,0.1);
  \node at (0,0)[above=2pt] {\footnotesize $ij$};
  \draw[quark2] (0,0) -- (1,0) node[pos=0.85, above=1pt] {\footnotesize $f$};
  \end{tikzpicture}
  &
  $\quad m_i \delta^{fj}$
  \\[3ex]
  $\op_3^i$      &
  \begin{tikzpicture}[scale=1, baseline=(current bounding box.center)]
  \draw[quark1] (-1,0) -- (0,0) node[pos=0.15, above=1pt] {\footnotesize $f$};
  \fill [black] (-0.1,-0.1) rectangle (0.1,0.1);
  \node at (0,0)[above=2pt] {\footnotesize $i$};
  \draw[quark2] (0,0) -- (1,0) node[pos=0.85, above=1pt] {\footnotesize $f$};
  \draw[->] (-0.5,-0.3) -- (0.5,-0.3) node[pos=0.5, below=0.5pt] {\footnotesize $p$};
  \end{tikzpicture}
  &
  $\quad (\slashed{p} - m_i) \delta^{fi}$
  \\[4ex]
  $\op_4$        &
  \begin{tikzpicture}[scale=1, baseline=(current bounding box.center)]
  \draw[gluon] (-1,0) -- (-0.1,0) node[pos=0.2, above=1pt] {\footnotesize $a,\alpha$};
  \fill [black] (-0.1,-0.1) rectangle (0.1,0.1);
  \draw[gluon] (0.1,0) -- (1,0) node[pos=0.8, above=1pt] {\footnotesize $b,\beta$};
  \draw[->] (-0.9,-0.25) -- (-0.4,-0.25) node[pos=0.5, below=0.5pt] {\footnotesize $p_1$};
  \draw[->] (0.4,-0.25) -- (0.9,-0.25) node[pos=0.5, below=0.5pt] {\footnotesize $p_2$};
\end{tikzpicture}
 &
 $\quad \delta^{ab} \big[ p_1^\alpha p_1^\beta + p_2^\alpha p_2^\beta  - g^{\alpha\beta}\big( p_1^2+p_2^2\big) \big]$
\\\bottomrule
 \end{tabular}
 \caption{Feynman rules for the operators at $\ord{g^0}$ relevant for the projection operators discussed in \subsec{projection}.
 \label{tab:FeynmanRulesOp}}
\end{table}
\endgroup

The Wilson coefficients can be extracted by means of projection operators~\cite{Gorishnii:1983su,Gorishnii:1986gn,Surguladze:1990sp} acting on both sides of the OPE in \eq{OPE}.
At this point we still choose the currents in \eq{correlatordef} to be bare.
The associated time-ordered product of these unrenormalized currents is denoted by $T_{(\mu\nu)}^{J,\bare}$.
Also the operators on the RHS of \eq{OPE} are considered bare at this stage.
In this way we obtain the bare Wilson coefficients $\C^{J, \bare}_n$.
The renormalization of the currents as well as of the operators
(and thus the Wilson coefficients)
is discussed in \sec{renormalization}.
The projection operators are defined by off-shell matrix elements
involving external one-particle or vacuum states and derivatives w.r.t.\ quark masses and external momenta together with a prescription to finally perform the zero-mass and zero-momentum limit, as explained below.
This implies that at intermediate steps also some of the gauge-dependent operator Wilson coefficients can appear.
The projectors are defined such that the operator matrix elements occuring on the RHS of \eq{OPE} only contribute at tree-level, i.e.\ their loop corrections are scaleless and vanish in dimensional regularization.
This allows to construct linear combinations of projection operators that determine the Wilson coefficients $\C^{J,\bare}_{1(,\mu\nu)}$,  $\C_{2(,\mu\nu)}^{J,\bare,ij}$
and to identify $\C_{6(,\mu\nu)}^{J,\bare,ijk}$ and $\C_{6(,\mu\nu)}^{J,\bare,ij}$ from the quark mass dependence
of the vacuum expectation value of the correlators as described in the following.

\begin{figure}[t]
  \begin{center}
    \includegraphics[width=0.316 \textwidth]{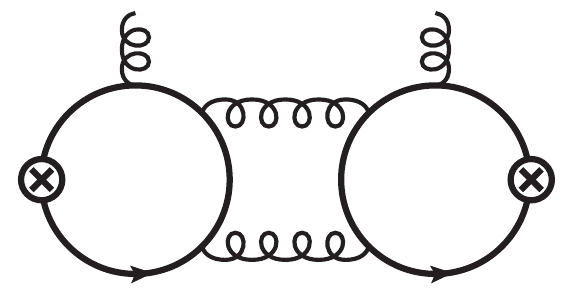}%
    \put(-75,75){(a)}
    \quad
    \raisebox{+1 pt}{\includegraphics[width=0.317 \textwidth]{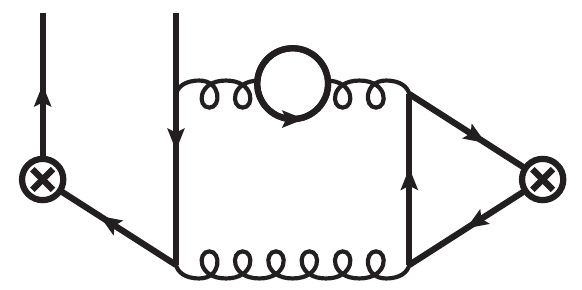}}%
    \put(-75,75){(b)}
    \quad
    \raisebox{-9 pt}{\includegraphics[width=0.315\textwidth]{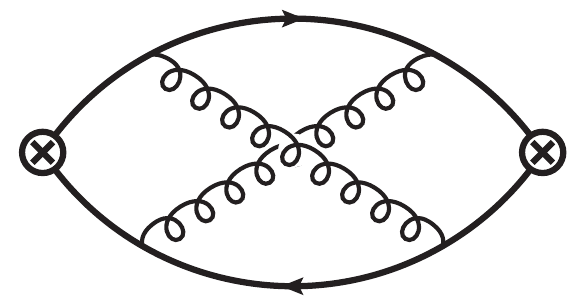}}%
    \put(-75,75){(c)}
  \end{center}
  \caption{
    Three-loop sample diagrams contributing to the Wilson coefficients of the dimension-four operators in the OPE of the current correlators \eq{correlatordef} via the projectors in eqs.~\eqref{eq:P1},~\eqref{eq:P23}, and \eqref{eq:P3}.
    Diagram~(a) is non-zero only if both external currents (symbolized by $\otimes$) have a flavor-singlet component according to \eq{fierz}.
    In that case it contributes to the Wilson coefficient $\C_{1,\mu \nu}^V$.
    Without a mass derivative acting on the quark triangle on the right,
    diagram~(b) is non-vanishing only if the current on the right has a flavor singlet component. With a mass derivative acting on the right triangle, non-vanishing contributions can arise for singlet and diagonal non-singlet currents.
    This diagram contributes to the coefficients
    $\C_{2,(\mu \nu)}^{J,ij}$ with $J=A,S,P$.
    Diagram~(c) can contribute to $\C_{6(,\mu\nu)}^{J,ijk}$ and $\C_{6(,\mu\nu)}^{J,ij}$ for $J=V,A,S,P$ in singlet--singlet, non-singlet--non-singlet, as well as singlet--non-singlet flavor configurations of the two currents.
    \label{fig:3loopdiags}
  }
\end{figure}

For the determination of $\C^{J,\bare}_{1(,\mu\nu)}$ we consider a matrix element with one incoming and one outgoing gluon. An external gluon state with momentum $p$, polarization $r$, and color $c$ we denote as $| \mathrm{g}(p,r,c) \rangle$. The corresponding projection operator then reads~\cite{HarlanderDiss}
\begin{align}
\mathrm{P}_{1}\big[T_{(\mu\nu)}^J \big] =&{} \lOP{\substack{p_k \to 0 \\ m_k \to 0}} \frac{ g^{\rho \sigma} }{4 d (d-1) N_A} \,  \frac{\partial^2}{\partial p_1^{\rho} \, \partial p_2^{\sigma}}  \nn\\
&{} \qquad \sum_{r_1,r_2=0}^3 \sum_{c=1}^{N_A} \varepsilon^*_\tau(p_1,r_1) \varepsilon^\tau(p_2,r_2)
 \big\langle \mathrm{g}(p_2,r_2,c) \; \big| T^J_{(\mu\nu)} \big| \mathrm{g}(p_1,r_1,c) \big\rangle
 \notag\\
 =&{}\lOP{\substack{p_k \to 0 \\ m_k \to 0}} \frac{ g^{\rho \sigma} g^{\alpha \beta} \delta^{ab}}{4 d (d-1) N_A} \,  \frac{\partial^2}{\partial p_1^{\rho} \, \partial p_2^{\sigma}} \, \Bigg(
 \begin{tikzpicture}[scale=1, baseline=(current bounding box.center)]
  \draw[gluon] (-1.5,0) -- (-0.5,0) node[pos=0.2, above=1pt] {\footnotesize $a,\alpha$};
  \draw[thick, rounded corners=10pt,fill=black!13] (-0.5,-0.5) rectangle (0.5,0.5);
  \node at (0,0) {\footnotesize $T^J_{(\mu\nu)}$};
  \draw[gluon] (0.5,0) -- (1.5,0) node[pos=0.8, above=1pt] {\footnotesize $b,\beta$};
  \draw[->] (-1.4,-0.2) -- (-0.8,-0.2) node[pos=0.5, below=0.5pt] {\footnotesize $p_1$};
  \draw[->] (0.8,-0.2) -- (1.4,-0.2) node[pos=0.5, below=0.5pt] {\footnotesize $p_2$};
  \draw[thick,cross,fill=white] (0,0.5) circle (0.1);
  \draw[thick,cross,fill=white] (0,-0.5) circle (0.1);
\end{tikzpicture}
\Bigg)
\label{eq:P1}
\end{align}
with $d=4-2\eps$ in dimensional regularization.
Here and below it is understood and absolutely crucial, that the derivatives and the zero-mass/momentum limit are taken on the integrand level \textit{before} performing any loop integrations.
According to the method of regions~\cite{Beneke:1997zp,Smirnov:2002pj} this yields exclusively the contributions from the region where all loop-momenta are hard ($\gg m_i, \Lambda_\mathrm{QCD})$, i.e.\ precisely the contributions that are to be encoded by the Wilson coefficients of the OPE.
Note that this implies in particular that self-energy corrections on the external (here gluon) legs are scaleless and vanish.
We therefore only need to compute amputated Feynman graphs.
This reflects the fact that wavefunction renormalization of the fields in the $\op_n$ does not affect their Wilson coefficients.%
\footnote{In any case, wavefunction corrections factorize and cancel between the LHS and RHS of the projected OPE equation, even before taking the zero-mass and zero-momentum limit.} A sample diagram at the three loop order,  i.e.\ $\ord{\as^3}$, is shown in \fig{3loopdiags}(a).
The projection prescription shown in the first line of \eq{P1} removes the polarization vectors of the two external gluon legs
in the off-shell matrix element.
In that expression the four polarization vectors (two from the external legs, two from the projector definition) are replaced by the corresponding polarization sums, $\sum_{r_i} \varepsilon^*_\mu(p_i,r_i) \varepsilon_\nu(p_i,r_i) \to - g_{\mu\nu}$ for $i=1,2$, to obtain the result in the last line.
The  sum of all amputated diagrams for the current correlator where the polarization vectors of the two external gluons are omitted is symbolized by the little graph in the last line of \eq{P1}. Similar prescriptions are also employed for the projectors discussed in the following.

From the Feynman rules for the operators in \tab{FeynmanRulesOp} we see that $\mathrm{P}_1$ exclusively projects onto $\C^{J,\bare}_{1(,\mu\nu)}$ on the RHS of the OPE:
\begin{align}
 \mathrm{P}_{1}[T_{(\mu\nu)}^{J,\bare}]
 =
 \frac{\C^{J,\bare}_{1(,\mu\nu)}}{-q^2} \;.
 \label{eq:C1bare}
\end{align}
It is important that the external gluons in $\mathrm{P}_1$ have distinct momenta $p_1 \neq p_2$ prior to taking the derivatives, since otherwise $\mathrm{P}_1$ would also project onto $\op_4$.
One can also use $\mathrm{P}_1$ with $p_1 = p_2$ and remove the admixture from $\op_4$ by means of additional projection operators involving external ghost states, see \rcites{Surguladze:1990sp,HarlanderDiss}. We have also implemented that approach as a cross check and found perfect agreement with our result from \eq{C1bare}.

The projection onto the coefficient $\C_{2(,\mu\nu)}^{J,\bare,ij}$ proceeds in a similar fashion, but is a bit more complicated.
Instead of external gluons we now consider external quark states $| \textrm{q}(i,p,s,c) \rangle$ with flavor $i$, momentum $p$, spin $s$, and color $c$.
Due to the presence of $\op_3^i$ two different projection operators are needed to single out $\C^{J,ij}_{2,(\mu\nu)}$.  Following \rcites{Surguladze:1990sp,HarlanderDiss} we can define them as
\begin{align}
\mathrm{P}_{23}^{ij}[T^J_{(\mu\nu)}] =&{} \lOP{p \to 0} \, \lOP{m_k \to 0} \, \frac{1}{4 N_c} \, \frac{\partial}{\partial m_{i}} \sum_{s=\pm} \sum_{c=1}^{N_c} \frac{1}{m_j}
\big\langle \textrm{q}(j,p,s,c) \big| T^J_{(\mu\nu)} \big| \textrm{q}(j,p,s,c) \big\rangle
\nn\\
=&{} \lOP{p \to 0} \, \lOP{m_k \to 0}  \, \frac{1}{4 N_c} \, \frac{\partial}{\partial m_{i}} \, \tr \Bigg[
\begin{tikzpicture}[scale=1, baseline=(current bounding box.center)]
  \draw[quark1] (-1.5,0) -- (-0.5,0) node[pos=0.1, above=1pt] {\footnotesize $j$};
  \draw[thick, rounded corners=10pt,fill=black!13] (-0.5,-0.5) rectangle (0.5,0.5);
  \node at (0,0) {\footnotesize $T^J_{(\mu\nu)}$};
  \draw[quark2] (0.5,0) -- (1.5,0) node[pos=0.9, above=1pt] {\footnotesize $j$};
  \draw[->] (-1.4,-0.2) -- (-0.8,-0.2) node[pos=0.5, below=0.5pt] {\footnotesize $p$};
  \draw[->] (0.8,-0.2) -- (1.4,-0.2) node[pos=0.5, below=0.5pt] {\footnotesize $p$};
  \draw[thick,cross,fill=white] (0,0.5) circle (0.1);
  \draw[thick,cross,fill=white] (0,-0.5) circle (0.1);
\end{tikzpicture}
\Bigg]\,,
\label{eq:P23}
\\
\mathrm{P}_3^i[T^J_{(\mu\nu)}] =&{} \lOP{p \to 0} \, \lOP{m_k \to 0} \, \frac{g^{\rho \sigma}}{8 d N_c}\,  \frac{\partial^2}{\partial p^\rho \, \partial p^\sigma} \sum_{s=\pm} \sum_{c=1}^{N_c}
\big\langle \textrm{q}(i,p,s,c) \big| T^J_{(\mu\nu)} \big| \textrm{q}(i,p,s,c) \big\rangle
\nn\\
=&{} \lOP{p \to 0} \, \lOP{m_k \to 0} \, \frac{1}{4 d N_c}\,  \frac{\partial}{\partial p^\rho } \, \tr \Bigg[ \gamma^\rho \times
\begin{tikzpicture}[scale=1, baseline=(current bounding box.center)]
  \draw[quark1] (-1.5,0) -- (-0.5,0) node[pos=0.1, above=1pt] {\footnotesize $i$};
  \draw[thick, rounded corners=10pt,fill=black!13] (-0.5,-0.5) rectangle (0.5,0.5);
  \node at (0,0) {\footnotesize $T^J_{(\mu\nu)}$};
  \draw[quark2] (0.5,0) -- (1.5,0) node[pos=0.9, above=1pt] {\footnotesize $i$};
  \draw[->] (-1.4,-0.2) -- (-0.8,-0.2) node[pos=0.5, below=0.5pt] {\footnotesize $p$};
  \draw[->] (0.8,-0.2) -- (1.4,-0.2) node[pos=0.5, below=0.5pt] {\footnotesize $p$};
  \draw[thick,cross,fill=white] (0,0.5) circle (0.1);
  \draw[thick,cross,fill=white] (0,-0.5) circle (0.1);
\end{tikzpicture}
\Bigg] \,.
\label{eq:P3}
\end{align}
The traces in the second lines of \eqs{P23}{P3} are taken over Dirac and color indices in the fermion representation and arise from the spin and color sums indicated in the first lines, employing $\sum_s u(i,p,s) \bar{u}(i,p,s) = \slashed{p}+m_i$. In analogy to \eq{P1}, the little graphs in the second lines symbolize the sum of amputated off-shell diagrams without external spinors for the quark legs. Note that a factor of $2$ arises in the second equality of \eq{P3} from the derivative product rule acting on the spin sum
and the amputated diagrams.
All terms that vanish due to the derivatives and the zero mass/momentum limit are already dropped.
Applying both projectors on the OPE of the current correlators we obtain
\begin{align}
 \mathrm{P}_{23}^{ij}[ T_{(\mu\nu)}^{J,\bare} ]{}&
 =
 \frac{\C_{2(,\mu\nu)}^{J,\bare, ij}}{-q^2} - \frac{\delta_{ij} \, \C_{3(,\mu\nu)}^{J,\bare,i}}{-q^2} \;,
 \\
 \mathrm{P}_3^i[ T_{(\mu\nu)}^{J,\bare} ]&{}
 =
 \frac{\C_{3(,\mu\nu)}^{J,\bare,i}}{-q^2}\;.
\end{align}
This allows us to compute the bare Wilson coefficient as
\begin{align}
  \C_{2(,\mu\nu)}^{J,\bare,ij}
  = -q^2
  \Bigl( \mathrm{P}_{23}^{ij}[ T_{(\mu\nu)}^{J,\bare} ]
  +\delta_{ij}
   \mathrm{P}_3^i[ T_{(\mu\nu)}^{J,\bare} ] \Bigr)
  \label{eq:C2bare}
\end{align}
to $\ord{\as^3}$ from Feynman diagrams with up-to three loops like e.g.\ the one in \fig{3loopdiags}(b).
Note that the operator $\op_3^i$ vanishes by virtue of the full classical equation of motion of the quark field and thus does not contribute to physical S-matrix elements. Its Wilson coefficient $\C_{3(,\mu\nu)}^{J,i}$ may therefore be gauge-dependent.
Indeed we find that starting from three loops $\C_{3(,\mu\nu)}^{J,i}$ explicitly depends on the gauge parameter in $R_\xi$-gauge.

To obtain the bare Wilson coefficients of the mass correction and identity operators we consider the vacuum expectation value of $T_{(\mu\nu)}^{J,\bare}$ and expand \text{before} doing the loop integrations in the quark masses up to and including the fourth power.
In \fig{3loopdiags}(c) we show a typical three-loop Feynman diagram.
The bare Wilson coefficients can then be directly read off
as the coefficients of the resulting polynomial in the quark masses:
\begin{align}
 \big\langle 0 \big| T_{(\mu\nu)}^{J,\bare}  \big| 0 \big\rangle
  ={}&
  -q^2 \,\C_{0(,\mu\nu)}^{J,\bare}
  \,+\,  \sum_{\substack{ ij \\ i \neq j}} \C_{m^2(,\mu\nu)}^{J,\bare,ij}  \, m_i m_j
  \,+\,  \sum_{i} \C_{m^2(,\mu\nu)}^{J,\bare,i}  \, m_i^2
 \nn\\
 &+\,  \sum_{\substack{ ijk \\ i \neq j}} \frac{\C_{6(,\mu\nu)}^{J,\bare,ijk} }{-q^2} \, m_i m_j m_k^2
 \,+\,  \sum_{ ij} \frac{\C_{6(,\mu\nu)}^{J,\bare,ij} }{-q^2} \, m_i^2 m_j^2
 \,+\, \mathcal{O}\biggl(\frac{m^6}{q^4}\biggr)\,.
 \label{eq:massexp}
\end{align}
In this work we are interested only in the mass corrections of power four, i.e.\ the last two terms of \eq{massexp}, for which we carry out the integrations up to the three-loop level and thus obtain results to $\ord{\as^2}$.
State-of-the-art results for the (UV-divergent) $\ord{m^2}$ coefficients and the (UV-divergent) identity coefficient, i.e.\ the massless vacuum polarization, can be found in \rcites{HarlanderDiss,Chetyrkin:1998ix,Harlander:1997kw,Harlander:1997xa,Chetyrkin:1997qi,Gorishnii:1986pz,Baikov:2008jh,Maier:2011jd}.
We also computed $\C_{0(,\mu\nu)}^{J,\bare}$ to three loops, compared the $\msb$ subtracted $\C_{0,\mu\nu}^{V}$ as well as  $\C_{0}^{S}$ to the expressions given in \rcites{HarlanderDiss,Baikov:2012zm} as a cross check of our computational setup and found perfect agreement.
Finally, we checked that our implementation of the Larin scheme to deal with the axial-vector and pseudo-scalar currents, see \subsec{Larin}, correctly removes the associated sub-divergences in $\C_{0,\mu\nu}^{A,\bare}$ and  $\C_{0}^{P,\bare}$.

\subsection{Calculation of the Feynman diagrams}
\label{subsec:calcdiags}

We generate the relevant Feynman graphs with \texttt{qgraf}~\cite{Nogueira:1991ex}. Next, we assign the individual diagrams according to their propagator structure to suitably defined integral families.
For this mapping we neglect the quark masses and the external momenta of the quark and gluon legs, since, according to the projections described in \subsec{projection}, they are set to zero before performing the loop integrations.
Hence, we only need to consider two-point integral families. This mapping procedure is automated in the (still) private \texttt{Looping} code~\cite{Looping} by one of the authors.

We process the integrands of the individual diagrams using \texttt{FORM}~\cite{Ruijl:2017dtg} as follows.
We first substitute the Feynman rules and compute the color factor of each diagram with the \texttt{color.h} package~\cite{vanRitbergen:1998pn}
in terms of the
group invariants and constants for a generic (compact semi-simple) gauge group.
For the case of SU($N_c$) with $T_F=1/2$ the latter are
\begin{align}
N_A = N_c^2 - 1 \,, & & C_F = \frac{N_c^2-1}{2 N_c} \,, & & C_A = N_c \,, & & \dFFF = \frac{(N_c^2 - 4)(N_c^2 - 1)}{16 N_c} \,.
\end{align}
In the case of vector and axial-vector currents we then proceed by applying additional projectors to the Feynman diagrams in order to extract the coefficients of the two Lorentz structures in \eq{WC-TL}.
Next, we expand the quark propagators in the quark masses.
This is necessary for the computation of the mass corrections and simplifies the computation of the projector $\mathrm{P}_{23}$. In the case of the projector $\mathrm{P}_1$ and $\mathrm{P}_{3}$ we can directly set the quark masses to zero.
In our flavor space setup involving the current density matrices $\rho_1,\rho_2$ (and flavor traces, see below), it is convenient to consider the quark propagators as a diagonal matrix in flavor space. To this end we define the diagonal quark mass matrix $M=\textrm{diag}(m_u,m_d,m_s,\dots)$
and employ the expanded expression
\begin{align}
\label{eq:propexp}
\frac{\ri}{\slashed{p}\otimes \bbid_{n_f \times n_f} - \bbid_s \otimes M} = \frac{\ri}{p^2} \big(\slashed{p}\otimes \bbid_{n_f \times n_f} + \bbid_s \otimes M \big) \left( \bbid_s \otimes \sum_{\ell=0}^\infty \left[ \frac{M^2}{p^2} \right]^\ell \right)\,,
\end{align}
for the quark propagator (truncated at order $\mathcal{O}(M^4)$ for the mass corrections and at order $\mathcal{O}(M)$ for the projector $\mathrm{P}_{23}$) prior to taking the derivatives w.r.t.\ masses and momenta
for the projectors.
Here $\bbid_s$ and $\bbid_{n_f \times n_f}$ are the identity matrices in spinor and flavor space, respectively.
After the masses and momenta are set to zero we compute the traces of Dirac matrices.
Finally we express each diagram as a linear combination of scalar loop integrals, which are members of the two-point integral family
the diagram was assigned to.
All necessary input files for carrying out the described steps of processing the integrand with \texttt{FORM} are generated automatically by \texttt{Looping} upon a minimal input of the user, in particular providing the Feynman rules and the projector definitions.

The final step in obtaining the results of the projected diagrams is the integration by parts (IBP) reduction~\cite{Tkachov:1981wb,Chetyrkin:1981qh}, where the scalar loop integrals are related to a small set of well-known three-loop two-point master integrals, see for instance \cite{Chetyrkin:1981qh,Chetyrkin:1980pr}.
We took the explicit expressions for the master integral from earlier work of two of the authors~\cite{Bruser:2018rad}.
For the IBP reduction we use \texttt{FIRE6}~\cite{Smirnov:2019qkx} in combination with \texttt{LiteRed}~\cite{Lee:2012cn,Lee:2013mka}.
We performed all calculations of the projected diagrams in general covariant $R_\xi$--gauge.
We verified that in the expressions for the Wilson coefficients $\C_{1(,\mu\nu)}^{J,\bare}$, $ \C_{2(,\mu\nu)}^{J,\bare,ij}$, $\C_{6(,\mu\nu)}^{J,\bare,ijk}$, $\C_{6(,\mu\nu)}^{J,\bare,ij}$ in terms of the linearly independent master integrals the dependence on the gauge parameter cancels out.
This serves as a strong check of our calculations.

\subsection{Treatment of $\gamma_5$ }
\label{subsec:Larin}

In order to consistently utilize dimensional regularization ($d=4-2\eps$)  in our computations involving the axial-vector and pseudo-scalar currents that are defined using the intrinsically four-dimensional Dirac matrix $\gamma_5$, see \eq{currentsdef},
we proceed as follows.
The corresponding Feynman diagrams contain exactly two $\gamma_5$ matrices.
We distinguish two-types of graphs.
In so-called non-singlet diagrams,
where the two currents are connected by at least one quark line,
both $\gamma_5$ matrices are within the same Dirac trace. An example is shown in \fig{3loopdiags}(c).
In contrast, singlet diagrams,
where the currents are not connected by any quark line,
have two traces with one $\gamma_5$ each, as e.g.\ in \fig{3loopdiags}(a) and~(b).
Note that in principle the diagrams can have additional traces from closed fermion loops like the one dressing the upper gluon line in \fig{3loopdiags}(b).
Those traces are however free of $\gamma_5$ matrices.
For the computation of the singlet diagrams we use the Larin scheme.
In the case of non-singlet diagrams we can employ the Larin scheme as well as naive dimensional regularization.
We briefly review these two  $\gamma_5$ schemes in the following.

Naive dimensional regularization~\cite{Bardeen:1972vi,Chanowitz:1979zu}
relies on promoting the anti-commutativity of $\gamma_5$ with all other $\gamma$ matrices to $d$ dimensions.
In the case of our non-singlet diagrams we first anticommute the two $\gamma_5$ matrices inside the trace such that they are adjacent to each other. We then use the identity $(\gamma_5)^2 = \bbid$ and simply compute the resulting trace without $\gamma_5$ in $d$ dimensions. This strategy is of course not applicable to the singlet diagrams.

Using the Larin scheme~\cite{Larin:1991tj,Larin:1993tq}, which is conceptually based on \rcite{tHooft:1972tcz},
for our computation of the bare Wilson coefficients we actually do not have to distinguish between singlet and non-singlet diagrams.
 In the Larin scheme $\gamma_5$ is replaced in the external currents using the following prescriptions:
 \begin{align}
  \gamma_\mu \gamma_5 \to \frac{i}{3!} \varepsilon_{\mu\nu\rho\sigma} \gamma^\nu \gamma^\rho \gamma^\sigma & \qquad \text{(axial-vector current)} \,,\nn \\
  \gamma_5 \to \frac{i}{4!} \varepsilon_{\mu\nu\rho\sigma} \gamma^\mu  \gamma^\nu \gamma^\rho \gamma^\sigma & \qquad \text{(pseudo-scalar current)} \,. \label{eq:Larin_prescriptions}
 \end{align}
 In our calculation this leads to a product of two Levi--Civita--tensors, that we rewrite as
 \begin{align}
  \varepsilon_{\mu_1 \mu_2 \mu_3 \mu_4} \varepsilon^{\nu_1 \nu_2 \nu_3 \nu_4}  = \textrm{det}
  \begin{pmatrix}
  g_{\mu_1}^{\nu_1} & \hdots & g_{\mu_4}^{\nu_1} \\
  \vdots & \ddots & \vdots \\
  g_{\mu_1}^{\nu_4} & \hdots & g_{\mu_4}^{\nu_4}
  \end{pmatrix}
  .
  \label{eq:Larin_LeviCivita}
 \end{align}
Although the Levi-Civita tensor is a strictly four--dimensional quantity, the metric tensors $g^{\mu\nu}$ on the RHS of this equation are eventually evaluated in $d$ dimensions.
Note that using \eqs{Larin_prescriptions}{Larin_LeviCivita} effectively leads to a non-anticommuting $\gamma_5$~\cite{tHooft:1972tcz} and spoils the chiral Ward identities including the all-order axial-anomaly equation.
These (anomalous) Ward identities are eventually restored via finite current renormalization factors~\cite{Larin:1991tj,Larin:1993tq,Ahmed:2021spj}, see \sec{renormalization}.
For the practical implementation of the Larin scheme in our code we follow \rcite{Moch:2015usa}.

We performed the calculation of the non-singlet diagrams both in the Larin scheme and using naive dimensional regularization. The results are in perfect agreement after (Larin scheme) renormalization of the external currents which serves as a stringent test of our implementation.

\section{Renormalization of the Wilson coefficients}
\label{sec:renormalization}

In this section we renormalize the OPE operators in \eq{Ops} and the two external correlator currents in the $\msb$ scheme, which also renders the OPE Wilson coefficients UV and IR finite, i.e.\ (fully) renormalized.
As required by the renormalization of the axial-vector current in the Larin scheme~\cite{Larin:1991tj,Larin:1993tq} we explicitly distinguish SU($n_f$) singlet and non-singlet currents, generically defined as
\begin{align}
  j^J_{\rs} = \bar{q} \, \Gamma^J \, \bbid \, q \,, \qquad\qquad
  j^{J,a}_{\rns} =  \bar{q} \, \Gamma^J \, t^a \, q \,,
  \label{eq:singnonsingcurrent}
\end{align}
where $\Gamma^J$ represents the Dirac structure of the current $(J = V, A, S, P)$ according to \eq{currentsdef} (and Lorentz indices are suppressed).
Any current with arbitrary flavor content can be decomposed into a linear combination of singlet and non-singlet currents via the decomposition in \eq{fierz}.
The flavor singlet and non-singlet currents are renormalized as
\begin{align}
  j^{J, \mathrm{bare}}_{\rs} =
  Z^J_\rs \, j^{J}_{\rs} \,,
  \qquad\qquad
  j^{J, \mathrm{bare}}_{\rns} =
  Z^J_\rns \, j^{J}_{\rns} \,,
\end{align}
respectively, with the  renormalization factors
\begin{align}
  Z^V_\rs ={}& Z^V_\rns = 1\,, \qquad   Z^S_\rs = Z^S_\rns = Z_m \,, \nn\\
  Z^A_\rs ={}& Z_\msb^\rs \, Z_5^\rs\,, \qquad Z^A_\rns = Z_\msb^\rns \, Z_5^\rns\,, \nn\\
  Z^P_\rs ={}& Z^P_\rns = Z_\msb^P \, Z_5^P \,,
  \label{eq:ZfactorsCurr}
\end{align}
where $Z_5^\rs$, $Z_5^\rns$, and $Z_5^P$ represent the finite renormalization factors for the axial-vector and pseudo-scalar currents in the Larin scheme, and all other $Z$-factors are pure $\msb$ renormalization constants. They are specified in more detail at the end of this section and given explicitly in \app{renconst}.
The renormalization of the UV sub-divergences in the bare expressions of the Wilson coefficients $\C^{J,\bare}_n$, obtained through the calculation described in \sec{calculation}, is performed by expressing the bare coupling $\as^\bare = (g^\bare)^2/(4\pi) = \tilde{\mu}^{2 \eps} Z_\alpha \as(\mu)$ in terms of the $\msb$ renormalized $\as$
with $\tilde{\mu}^2= \mathrm{e}^{\gamma_\mathrm{E}}\mu^2/(4\pi)$
and incorporating the appropriate $Z$-factors according to \eq{ZfactorsCurr} for the renormalization of the two external currents.
We thus have for the UV (sub-)renormalized Wilson coefficients
\begin{align}
 \C_n^{J,\sub}(\as) &= \Bigr[ \C_n^{J,\bare}(\as^\bare \!=\!  \tilde{\mu}^{2 \eps}  Z_\alpha \as)  \Bigr]^{\tr[\rho_i]\to Z_\rs^J \tr[\rho_i]}_{\tr[\rho_i t^a] \to Z_\rns^J \tr[\rho_i t^a]}\,,
 \label{eq:Csub}
\end{align}
where the decomposition of the current flavor matrices $\rho_i$ ($i=1,2$) into singlet ($\rs$) and non-singlet ($\rns$) contributions using \eq{fierz} is implied, and the implementation of the current renormalization factors is made explicit at the level of the flavor matrices.

With \eq{Csub} we have performed  the renormalization of the currents and the coupling, but not of the OPE operators in \eq{Ops},
so that the Wilson coefficients are still IR divergent. They are rendered IR finite by renormalizing the bare OPE operators.
As the operators (and thus the associated Wilson coeffcients) mix, the renormalization $Z$-factor for the dimension-four operators in \eq{Ops} is matrix-valued~\cite{Kluberg-Stern:1974iel,Spiridonov:1984br,Spiridonov:1988md,Chetyrkin:1994ex}.
To specify an explicit form of the $Z$-matrix, let us define the following vector of the physically relevant operators:
\begin{align}
  \vec{\op}^{\,\rT} \equiv \Bigl(
  \op_1,
  \bigl(\vec{\op}_2^\rd\bigr)^{\!\rT} ,
  \bigl(\vec{\op}_2^\rnd \bigr)^{\!\rT} ,
  \bigl(\vec{ \op}_6^\rd\bigr)^{\!\rT},
  \bigl(\vec{ \op}_6^{\rnd1}\bigr)^{\!\rT},
  \bigl(\vec{ \op}_6^{\rnd2} \bigr)^{\!\rT},
  \bigl(\vec{\op}_6^{\rnd3} \bigr)^{\!\rT}
  \Bigr) \,,
  \label{eq:opvec}
\end{align}
where
\begin{align}
  \bigl(\vec{\op}_2^\rd \bigr)^{\!\rT}
  &\equiv
  \bigl(
  \op_2^{uu},  \op_2^{dd}, \ldots
  \bigr) = \bigl(m_u \bar{u}u \,,\, \ldots \bigr),
  \nn\\
  \bigl(\vec{\op}_2^\rnd \bigr)^{\!\rT}
  &\equiv
  \bigl(
  \op_2^{ud},  \op_2^{us}, \ldots, \op_2^{du}, \op_2^{ds}, \op_2^{dc}, \ldots
  \bigr) = \bigl( m_u \bar{d}d \,,\,\ldots \bigr) \,, \nn\\
  \bigl(\vec{\op}_6^\rd \bigr)^{\!\rT}
  &\equiv
  \bigl(
  \op_6^{uu},  \op_6^{dd}, \ldots
  \bigr)=\bigl(m_u^4 \,,\, \ldots \bigr) \,,  \nn\\
  \bigl(\vec{ \op}_6^{\rnd1} \bigr)^{\!\rT}
  &\equiv
  \bigl(
  \op_6^{ud},  \op_6^{us}, \ldots, \op_6^{ds}, \op_6^{dc}, \ldots
  \bigr)   = \bigl( m_u^2 m_d^2\,,\,\ldots \bigr) \,, \nn\\
  \bigl(\vec{ \op}_6^{\rnd2} \bigr)^{\!\rT}
  &\equiv
  \bigl(
  \op_6^{udd},  \op_6^{uss}, \ldots, \op_6^{duu}, \op_6^{dss}, \op_6^{dcc}, \ldots
  \bigr) = \bigl( m_u m_d^3 \,,\,\ldots  \bigr) \,, \nn\\
  \bigl(\vec{ \op}_6^{\rnd3} \bigr)^{\!\rT}
  &\equiv
  \bigl(
  \op_6^{uds},  \op_6^{udc}, \ldots, \op_6^{dsc}, \op_6^{dsn}, \ldots
  \bigr) = \bigl( m_u m_d m_s^3 \,,\, \ldots \bigr) \,.
  \label{eq:opsubvec}
\end{align}
The operator spaces for $\vec{\op}_2^\rd$ and $\vec{\op}_6^\rd$ each have dimension $n_f$, for $\vec{\op}_2^\rnd$ and $\vec{ \op}_6^{\rnd2}$ each have dimension $n_f(n_f-1)$. The one for $\vec{ \op}_6^{\rnd1}$ has dimension $n_f(n_f-1)/2$ and the one for $\vec{ \op}_6^{\rnd3}$ has dimension $n_f(n_f-1)(n_f-2)/2$.
The relation of bare and renormalized composite operators reads
\begin{align}
  {\vec{\op}\,} = Z\, {\vec{\op}^{\,\bare}}\,,
  \label{eq:OZrenormalized}
\end{align}
where the renormalization $Z$ matrix has the form
\begin{align}
  Z =
  \begin{pmatrix}
    Z_{11} &  Z_{12} {\vec{1}\,}^\rT & 0 & Z_{16}^\rd  {\vec{1}\,}^\rT & Z_{16}^\rnd\,  {\vec{1}\,}^\rT & 0  & 0 \\
    0 & \bbid & 0 & Z_{26}^\rd \bbid &  Z_{26}^\rnd \Delta& 0 &  0 \\
    0 & 0 & \bbid & 0 & 0  & Z_{26}^\rd  \bbid + Z_{26}^\rnd \widetilde{\Delta}&  Z_{26}^\rnd  \bar{\Delta}\\
    0 & 0 & 0 & Z_{66}\bbid & 0 & 0 &  0 \\
    0 & 0 & 0 & 0  & Z_{66} \bbid& 0 &  0 \\
    0 & 0 & 0 & 0 & 0 & Z_{66} \bbid&  0 \\
    0 & 0 & 0 & 0 & 0 & 0  &  Z_{66} \bbid
  \end{pmatrix}\,.
  \label{eq:Zmatrix}
\end{align}
All fields, couplings and masses in ${\vec{\op}^{\,\bare}}$ are understood as bare quantities.
The off-diagonal matrices $\Delta$, $\bar{\Delta}$, and $\widetilde{\Delta}$, which generate mixing between different operator sub-spaces,
are defined by
\begin{align}
  \Delta \, \vec{ \op}_6^{\rnd1} & =
  \Bigl(m_u^2 \sum_{i\neq u}m_i^2, \, m_d^2 \sum_{i\neq d}m_i^2,\, \ldots \, \Bigr)^\rT\,,
  \nn\\
  \widetilde{\Delta}\, \vec{ \op}_6^{\rnd 2}  & =
  \Bigl( m_u^3 m_d, m_u^3 m_s ,\, \ldots, \, m_d^3 m_u, m_d^3 m_s,\, \ldots\, \Bigr)^\rT\,,
  \nn\\
  \bar{\Delta}\, \vec{ \op}_6^{\rnd 3}  & =
  \Bigl(m_u m_d \!\sum_{k\neq u,d}m_k^2, \, m_u m_s \!\sum_{k\neq u,s}m_k^2,\, \ldots,\, m_d m_u \!\sum_{k\neq u,d}m_k^2, \, m_d m_s \!\sum_{k\neq d,s}m_k^2,\, \ldots\, \Bigr)^\rT
  \,.
  \label{eq:Deltamat}
\end{align}

\begin{figure}[t]
  \begin{center}
    \includegraphics[width=0.16 \textwidth]{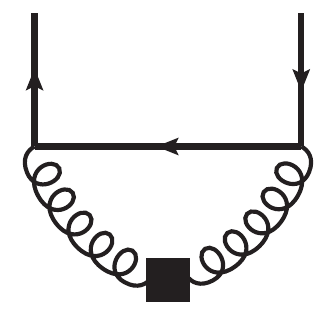}%
    \put(-72,70){(a)}
    \put(-40,-11){$\op_1$}
    \qquad
    \includegraphics[width=0.15 \textwidth]{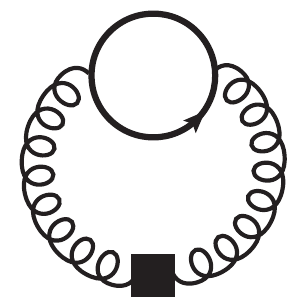}%
    \put(-70,70){(b)}
    \put(-38,-11){$\op_1$}
    \qquad
    \includegraphics[width=0.175 \textwidth]{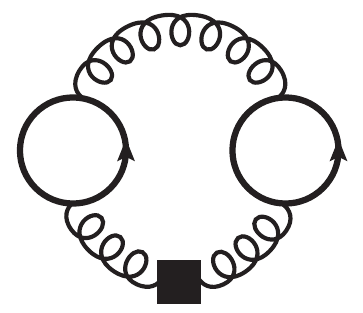}%
    \put(-80,70){(c)}
    \put(-44,-11){$\op_1$}
    \qquad
    \includegraphics[width=0.15 \textwidth]{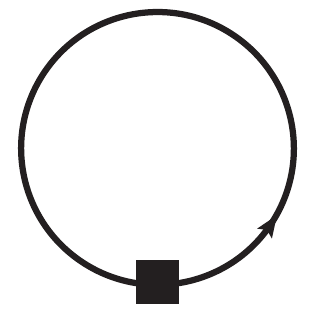}%
    \put(-70,70){(d)}
    \put(-38,-11){$\op_2^{ij}$}
    \qquad
    \includegraphics[width=0.15 \textwidth]{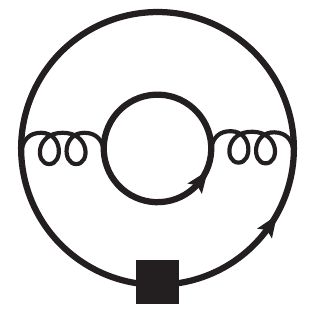}%
    \put(-70,70){(e)}
    \put(-38,-11){$\op_2^{ij}$}
  \end{center}
  \caption{
    The UV divergences of the diagrams (a), (b), (c), (d), and (e) determine the  $Z$-factors in the off-diagonal entries of the renormalization matrix $Z$ in \eq{Zmatrix}, i.e.\ $Z_{12}$, $Z_{16}^\rd$, $Z_{16}^\rnd$, $Z_{26}^\rd$, and $Z_{26}^\rnd$, at leading order in the strong coupling.
    Note that, because the  flavor of both quark loops may be equal, diagrams (c) and (e) also contribute to $Z_{16}^\rd$ and $Z_{26}^\rd$, but not to $Z_{16}^{\mathrm{d,ex}}=Z_{16}^{\rd} - Z_{16}^{\rnd}$ and $Z_{26}^{\mathrm{d,ex}}=Z_{26}^{\rd} - Z_{26}^{\rnd}$.
    \label{fig:Z16Z26}
  }
\end{figure}
Through \eq{OZrenormalized}
S-matrix elements or vacuum expectation values with insertions of the renormalized operators are finite.
We note that the different $Z$-factors shown in the matrix on the RHS of \eq{Zmatrix} are all number-valued and that the matrix character is fully captured by
the $\Delta$ matrices in \eq{Deltamat}, by ${\vec{1}} = (1,1,1,\ldots)^\rT$ and by the corresponding identity matrices $\bbid$. The origin of the $Z$-factors related to the operator mixing appearing in the off-diagonal blocks is illustrated diagrammatically in \fig{Z16Z26}.
They have the structure $Z_{12} = \ord{\as}$,
$Z_{16}^\rd = \ord{\as}$, $Z_{16}^\rnd = \ord{\as^2}$, $Z_{26}^\rd = \ord{\as^0}$, and $Z_{26}^\rnd  = \ord{\as^2}$.
The $Z$-factors in the diagonal blocks have the structure $Z_{11}=1 +\ord{\as}$ and  $Z_{66}=1+\ord{\as}$ and are related to the renormalization constants of the QCD Lagrangian in \eq{ZfactorsOps}.

In the vector notation corresponding to \eqs{opvec}{OZrenormalized}
the renormalized Wilson coefficients read
\begin{align}
   \vec{\C}^{J}(\as) = \bigl(Z^{-1}\bigr)^{\!\mathrm{T}} \, \vec{\C}^{J,\sub}(\as) \, .
   \label{eq:Crenvec}
\end{align}
We note that for the determination of the matrix $Z^{-1}$, due to the triangular structure of $Z$, only the inverses of the $Z$-factors in the diagonal blocks are required, but no inverse of any off-diagonal block.
In view of a compact presentation of the final results in \sec{results}, we define the shorthand notations
\begin{align}
  \CO_1^{J(,\sub)} &\equiv \C_1^{J(,\sub)} \op_1\,, \qquad
  \CO_2^{J(,\sub)} \equiv \sum_{ij} \C_2^{J(,\sub),ij} \op_2^{ij}\,,
  \nn\\
  \CO_6^{J(,\sub)} &\equiv \sum_{ijk} \C_6^{J(,\sub),ijk} \op_6^{ijk} +
  \sum_{ij} \C_6^{J(,\sub),ij} \op_6^{ij} \,
  \label{eq:CO}
\end{align}
for the sums of the renormalized operators $\op_n$ multiplied with their respective (still IR divergent) Wilson coefficients,
such that
\begin{align}
  \vec{\C}^{J,\sub} \cdot {\vec{\op}^{\,\bare}}
  = \vec{\C}^{J} \! \cdot \vec{\op}
  = \CO_1^J + \CO_2^J+\CO_6^J\,,
\end{align}
where here and in the following we suppress the argument $\alpha_s$ for brevity.

Using \eq{CO} we can now cast the matrix relation of \eq{Crenvec}
in the form of only three remarkably compact equations:
\begin{align}
  \CO_1^J ={}& Z_{11}^{-1} \, \CO_1^{J,\sub} \,, \label{eq:renCO1}\\
  \CO_2^J ={}& \CO_2^{J,\sub} -   Z_{11}^{-1}Z_{12}\,
  \,\C_1^{J,\sub}
  \sum_i \op_2^{ii}
  \,=\, \CO_2^{J,\sub} - Z_{11}^{-1}Z_{12}\,
  \,\C_1^{J,\sub}
  \tr[M \Q]
  \,,
  \label{eq:renCO2}\\
  \CO_6^J ={}& Z_{66}^{-1}\biggl\{ \CO_6^{J,\sub}
  \!+ Z_{11}^{-1}
  \Bigl[
  \bigl(Z_{12} Z_{26}^{\mathrm{d,ex}} \!- Z_{16}^{\mathrm{d,ex}} \bigr)  \tr\bigl[M^4\bigr]
  +  \bigl(Z_{12} Z_{26}^{\rnd} - Z_{16}^{\rnd} \bigr)  \tr\bigl[M^2\bigr]^2 \Bigr]
  \C_1^{J,\sub} \nn\\
   &\qquad - Z_{26}^{\mathrm{d,ex}}
   \Bigr[ \CO_2^{J,\sub}\Bigr]_{\bar{q}_i q_i \to m_i^3}
  - Z_{26}^{\rnd} \, \tr\bigl[M^2\bigr]
  \Bigr[ \CO_2^{J,\sub}\Bigr]_{\bar{q}_i q_i \to m_i}
  \biggr\}\,.
  \label{eq:renCO6}
\end{align}
In \eq{renCO2} we have used
the diagonal quark mass matrix $M = \mathrm{diag}(m_u, m_d, m_s, \ldots)$ from \eq{propexp}
and introduced the diagonal quark condensate (operator) matrix $\Rrc = \mathrm{diag}(\bar{u}u, \bar{d}d, \bar{s}s, \ldots)$. The trace $\tr[M \Q]$ which multiplies the term arising from the mixing with the gluon condensate, equals the linear combination $\sum_i \op_2^{ii}$ of renormalized operators $\op_2^{ij}$.
The renormalized operators simply act as polynomal placeholders to identify the expressions for the renormalized Wilson coefficients $\C_2^{J,ij}$.
The latter are then given by the sum of the terms multiplying each $\op_2^{ij}$ on the RHS.
In \eq{renCO6}, which entails the results for the renormalized Wilson coefficients  $\C_6^{J,ij}$ and $\C_6^{J,ijk}$,
we use the analogous notation. The replacements
$\bar{q}_i q_i \to m_i$ and $\bar{q}_i q_i \to m_i^3$ refer to the polynomial placeholders for the corresponding renormalized mass operators arising due to mixing.
S-matrix elements with insertions of the $\CO^J_n$ are finite.

The renormalization $Z$-factors appearing in \eqsm{renCO1}{renCO6} and \eq{Zmatrix} are given by~\cite{Kluberg-Stern:1974iel,Spiridonov:1984br,Spiridonov:1988md,Chetyrkin:1994ex}
\begin{align}
  Z_{11} &= \biggl(1 - \frac{\beta(\as)}{\eps} \biggr)^{\!\!-1}, &
  Z_{12} &= - \frac{4 \gamma_m(\as)}{\eps}\, Z_{11}\,, & Z_{66} &= Z_m^{-4}\,, \nn\\
  Z_{16}^{\mathrm{d,ex}} &=Z_{16}^{\rd} - Z_{16}^{\rnd} \,, &
  Z_{26}^{\mathrm{d,ex}} &=Z_{26}^{\rd} - Z_{26}^{\rnd}\,, \nn\\
  Z_{16}^{\rd} & = 4 Z_m^{-4} \as \frac{\partial}{\partial \as}\, Z_0^{\rd}\,,
  &
  Z_{16}^{\rnd}  &= 4 Z_m^{-4} \as \frac{\partial}{\partial \as}\,
  Z_0^{\rnd}\,, \nn\\
  Z_{26}^{\rd} & = -4 Z_m^{-4} Z_0^{\rd}\,, &
  Z_{26}^{\rnd} &= -4 Z_m^{-4} Z_0^{\rnd}\,.
  \label{eq:ZfactorsOps}
\end{align}
The renormalization constants $Z_0^{\rd}$, $Z_0^{\rnd}$, from which the  $Z$-factors in the last two lines of \eq{ZfactorsOps} can be derived, are related to the corresponding anomalous dimensions of the QCD vacuum energy density~\cite{Spiridonov:1988md,Chetyrkin:1994ex}
\begin{align}
  \gamma_0^{\rd (\rnd)} = \bigl(4 \gamma_m(\alpha_s) - \eps \bigr) Z_0^{\rd (\rnd)}  +  \bigl(\beta(\alpha_s) - \eps \bigr) \as \frac{\partial}{\partial \as}  Z_0^{\rd (\rnd)} \,.
  \label{eq:gam0}
\end{align}
The anomalous dimensions $\gamma_0^{\rd}$, $\gamma_0^{\rnd}$ are given to five-loop order in \rcite{Baikov:2018nzi}, and we
use these to extract $Z_0^{\rd}$ and $Z_0^{\rnd}$ to $\ord{\as^2}$, i.e.\ three loops, from \eq{gam0}.
We give their explicit expressions  in \app{renconst}.
The result for $Z_0^{\rd}$ can also be found in \rcite{Harlander:2020duo}.
The renormalization factors $Z_{11}$ and $Z_{12}$ in \eq{ZfactorsOps} are expressed in terms of the ($\eps$-independent part of the) anomalous dimension of $\as$
and the ($\eps$-independent) anomalous dimension of the quark masses~\cite{Kluberg-Stern:1974iel,Spiridonov:1984br}:
\begin{align}
  \mu \frac{\rd}{\rd \mu}\, \as = 2 \as \bigl[ -\eps + \beta(\as) \bigr]\,,
  \qquad
  \mu \frac{\rd}{\rd \mu}\, m_i  = 2 \gamma_m(\as)\, m_i \,.
\end{align}
The latter is in turn tied to the mass renormalization constant $Z_m = m^\bare/m$, and thus to  $Z_{66} = Z_m^{-4}$, as
\begin{align}
  \gamma_m(\as) = - \mu^2 \frac{\rd}{\rd \mu^2} \ln Z_m = \bigl[ \eps - \beta(\as) \bigr] \as \frac{\partial}{\partial \as}
  \ln Z_m\,.
\end{align}
We take the required expressions for $\beta(\as)$, $\gamma_m(\as)$, $Z_m$, and $Z_\alpha$ from \rcite{Chetyrkin:2017bjc} (see also \rcite{Luthe:2017ttc}).
Besides that, we need the divergent ($Z_\msb^\rs$, $Z_\msb^\rns$, $Z_\msb^P$) and finite current renormalization factors ($Z_5^\rs$, $Z_5^\rns$, $Z_5^P$) in the Larin scheme, see \eq{ZfactorsCurr}. These are found in \rcites{Larin:1991tj,Larin:1993tq,Ahmed:2021spj}.
For convenience of the reader, we collect all relevant renormalization constants in \app{renconst}.

\section{Results}
\label{sec:results}

The implementation of the operator renormalization through  \eqsm{renCO1}{renCO6} renders all Wilson coefficients (UV as well as IR) finite.
To this end we write the $\CO_n^{J,\sub}$ defined in \eq{CO} in the form
\begin{align}
    \CO_1^{J,\sub} ={}&
	\sum_{i_1, i_2, j_1, j_2}\!\! F^1_{i_1 i_2 j_1 j_2} \op_1\,,\\
	\CO_2^{J,\sub} ={}&
	\sum_{i_1, i_2, j_1, j_2}\!\! F^2_{i_1 i_2 j_1 j_2} M_{i_1 i_2} \mathrm{Q}_{j_1 j_2}  \;+\;
	\sum_{i_1, i_2} F^3_{i_1 i_2} (M \mathrm{Q})_{i_1 i_2} \,,
	\label{eq:CO2Fs} \\
	\CO_6^{J,\sub} ={}&   \sum_{i_1, i_2, j_1, j_2,k_1,k_2} \!\! F^4_{i_1 i_2 j_1 j_2,k_1,k_2} M_{i_1 i_2} M_{j_1 j_2}  (M^2)_{k_1 k_2}
	\;+\sum_{i_1, i_2} F^5_{i_1 i_2} (M^4)_{i_1 i_2}   \nn  \\
	&+\sum_{i_1, i_2, j_1, j_2} \!\! F^6_{i_1 i_2 j_1 j_2} (M^2)_{i_1 i_2} (M^2)_{j_1 j_2}
	\;+\sum_{i_1, i_2, j_1, j_2} \!\! F^7_{i_1 i_2 j_1 j_2} M_{i_1 i_2} (M^3)_{j_1 j_2}  \,,
	\label{eq:CO6Fs}
\end{align}
where, due to the Fierzing of $\rho_1$, $\rho_2$ in \eq{Csub} via \eq{fierz},
the flavor tensors $F^n$ are linear combinations of products of SU($n_f$) generators $t^a_{ij}$ and Kronecker deltas $\delta_{ij}$.
The coefficients of these linear combinations accordingly involve the flavor traces $\tr\bigl[\rho_1\bigr]$, $\tr\bigl[\rho_2\bigr]$, $\tr\bigl[\rho_1 t^a\bigr]$, and  $\tr\bigl[\rho_2 t^a\bigr]$.
Recall that $M = \mathrm{diag}(m_u, m_d, m_s, \ldots)$ and $\Rrc = \mathrm{diag}(\bar{u}u, \bar{d}d, \bar{s}s, \ldots)$.
The terms in the last line of \eq{renCO6} can be written in an analogous way, where the replacements $\bar{q}_i q_i \to m_i^3$ and $\bar{q}_i q_i \to m_i^3$ translate to $\Q \to M^3$ and $\Q \to M$, respectively.
Each $t^a_{ij}$ appearing in the $F^n$ tensors is accompanied by a corresponding factor  $\tr\bigl[\rho_1 t^a\bigr]$ or  $\tr\bigl[\rho_2 t^a\bigr]$ (with summation over the index $a$) originating from the decomposition in \eq{fierz}.
We can therefore eliminate each $t^a_{ij}$ by substituting it with a linear combination of $(\rho_{1})_{ij}$ or $(\rho_{2})_{ij}$ and $\delta_{ij}$ in the finite $\CO^J_n$ using once again \eq{fierz}.
We are thus left with $\CO^J_n$ expressions that can be written as linear combinations of products of flavor traces involving only $\rho_1$, $\rho_2$, $Q$, and powers of $M$.
After some straightforward algebraic manipulations exploiting e.g.\ the cyclicity of the flavor traces and that the matrices $M$ and $Q$ commute,
we obtain the following irreducible form of the results
\begin{align}
 \CO^{J}_1 ={}&  c_{1,1}^{J} \, \tr\bigl[ \rho_1 \bigr] \, \tr\bigl[ \rho_2 \bigr] \, \op_1 + c_{1,2}^{J} \, \tr\bigl[\rho_1 \rho_2 \bigr] \, \op_1 \,,
 \label{eq:CO1}
\\[2 ex]
\CO^{J}_2 ={}&  c_{2,1}^{J} \, \tr\bigl[\Rra\bigr] \, \tr\bigl[\Rrb\bigr] \, \tr\bigl[\Mm \Rrc\bigr]
 + c_{2,2}^{J} \, \tr\bigl[\Rra \Rrb\bigr] \, \tr\bigl[\Mm \Rrc\bigr]
 \notag\\
 & + c_{2,3}^{J} \Bigl( \tr\bigl[\Rra \Rrc\bigr] \, \tr\bigl[\Rrb \Mm\bigr] + \tr\bigl[\Rra \Mm\bigr] \, \tr\bigl[\Rrb \Rrc\bigr]\Bigr)
 \notag\\
 & + c_{2,4}^{J} \Bigl( \tr\bigl[\Rra\bigr] \, \tr\bigl[\Rrb \Mm \Rrc\bigr] + \tr\bigl[\Rrb\bigr] \, \tr\bigl[\Rra \Mm \Rrc\bigr]\Bigr)
 \notag\\
 & + c_{2,5}^{J} \Bigl( \tr\bigl[\Rra \Mm \Rrb \Rrc\bigr] + \tr\bigl[\Rrb \Mm \Rra \Rrc\bigr]\Bigr)
+ c_{2,6}^{J} \Bigl( \tr\bigl[\Rra \Rrb \Mm \Rrc\bigr] + \tr\bigl[\Rrb \Rra \Mm \Rrc\bigr]\Bigr) \,,
  \label{eq:CO2}
\\[2ex]
 \CO^{J}_6 ={}& \frac{N_c}{\pi^2} \biggl\{
c_{6,1}^{J} \Bigl( \tr\bigl[\Rra\bigr] \, \tr\bigl[\Rrb \Mmp{4}\bigr] + \tr\bigl[\Rrb\bigr] \, \tr\bigl[\Rra \Mmp{4}\bigr]\Bigr)
 \notag\\
 & + c_{6,2}^{J} \, \tr\bigl[\Mmp{4}\bigr] \, \tr\bigl[\Rra \Rrb\bigr]
  + c_{6,3}^{J} \, \tr\bigl[\Rra \Mmp{2}\bigr] \, \tr\bigl[\Rrb \Mmp{2}\bigr]
 \notag\\
 & + c_{6,4}^{J} \Bigl( \tr\bigl[\Rra \Mm\bigr] \, \tr\bigl[\Rrb \Mmp{3}\bigr] + \tr\bigl[\Rra \Mmp{3}\bigr] \, \tr\bigl[\Rrb \Mm\bigr]\Bigr)
 \notag\\
 & + c_{6,5}^{J} \Bigl( \tr\bigl[\Rra \Rrb \Mmp{4}\bigr] + \tr\bigl[\Rrb \Rra \Mmp{4}\bigr]\Bigr)
 \notag\\
 & + c_{6,6}^{J} \, \tr\bigl[\Mmp{2}\bigr] \Bigl( \tr\bigl[\Rra \Rrb \Mmp{2}\bigr] + \tr\bigl[\Rrb \Rra \Mmp{2}\bigr]\Bigr)
 + c_{6,7}^{J} \, \tr\bigl[\Rra \Mmp{2} \Rrb \Mmp{2}\bigr]
 \notag\\
 & + c_{6,8}^{J} \Bigl( \tr\bigl[\Rra \Mm \Rrb \Mmp{3}\bigr] + \tr\bigl[\Rrb \Mm \Rra \Mmp{3}\bigr]\Bigr)
+ c_{6,9}^{J} \, \tr\bigl[\Mmp{2}\bigr] \, \tr\bigl[\Rra \Mm \Rrb \Mm\bigr]
 \biggl\} \,.
\label{eq:CO6}
\end{align}
Note that for $J=V, A$ the two Lorentz structures $\CO_n^{J,T}$ and $\CO_n^{J,L}$
are implied according to \eq{WC-TL}, and the analogous decomposition also applies to the trace coefficients. The $c_{n,m}^J$ are given explicitly in \app{ResCoef}.
They can become non-vanishing for a particular current correlator, but not all of them are always non-zero for each current correlator. In $\CO^{J}_6$ additional trace structures with modified powers of the mass matrix $M$ may exist beyond three loop order due to additional quark loops.
All our results are also provided in the ancillary file.

QCD parity symmetry implies the invariance of the OPE Wilson coefficients under $\rho_1\leftrightarrow\rho_2$ exchange.
This requires some pairs of coefficients of the flavor traces in $\CO^J_2$ and  $\CO^J_6$ to be equal as made explicit in \eqs{CO2}{CO6}.
Charge conjugation invariance implies the symmetry under $\rho_i \leftrightarrow (-1)^n(\rho_i)^T$, with $n=1$ for vector currents and $n=0$ otherwise.
This constraint only affects the terms where $\rho_1$ and $\rho_2$ both appear within a single trace (such as $\tr\bigl[\Rra \Rrb \Mm \Rrc\bigr]$) and does not yield any constraints beyond those already provided by parity symmetry.

The representation in terms of flavor traces without the appearance of any generator $t^a$ allows  to diagrammatically interpret the structure of the $\CO^J_n$ via the  direct relation to flavor (quark) lines involving the two external currents, the quark mass derivatives and the quark condensates. For example, the diagram in \fig{3loopdiags}(a) contributes to the coefficient of $\tr\bigl[\Rra\bigr]\tr\bigl[\Rrb\bigr]$ in $\CO^V_1$.
The diagram in \fig{3loopdiags}(b)  contributes to the coefficient of
$\tr\bigl[\Rra \Rrc\bigr]\tr\bigl[\Rrb \Mm\bigr]$ in $\CO_2^{J}$ for the case that the left cross represents the $\rho_1$ current and the mass derivative (from the projector $\mathrm{P}_{23}^{ij}$) acts on the quark triangle involving $\rho_2$.%
\footnote{
For the diagram in \fig{3loopdiags}(b) and $J=S,P$ the contribution from the projector $\mathrm{P}_{3}^{i}$ vanishes because without a mass derivative the Dirac trace associated with the quark triangle on the right vanishes due to an odd number of $\gamma^\mu$ matrices involved.
The flavor trace of the quark bubble yields a factor $n_f$.
For $J=V$ this diagram yields no contribution at all because of a Dirac trace with an odd number of $\gamma_\mu$ (when the mass derivative acts on the quark triangle) or Furry's theorem upon adding the diagram with reversed fermion flow in the triangle.
For $J=A$ there is a non-zero contribution due to the axial anomaly and the mass derivative acting on the quark line on the left.}
The diagram in \fig{3loopdiags}(c) contributes to the coefficient of $\tr\bigl[\Rra \Rrb \Mmp{4}\bigr]$ in $\CO_6^J$  for all currents in the case that  the left cross again represents the $\rho_1$ current and the four mass derivatives act on the upper quark line.

The close relation of our representation to the underlying Feynman diagrams
permits in particular an easy and unambiguous identification of the contributions from certain diagram classes in the final results such as so-called singlet diagrams where the flavor lines of the two external currents are mutually disconnected. Diagrams of the latter type give rise to flavor traces involving $\rho_1$ and $\rho_2$ individually (potentially with additional insertions of the $M$ and $Q$ matrices), but not both of them, and are exclusively contained in the coefficients $c^J_{1,1}$, $c^J_{2,\{1,3,4\}}$ and $c^J_{6,\{1,3,4\}}$. Thus, if desired, the contributions of the singlet diagrams can be switched off by setting these coefficients to zero. Examples for singlet diagrams are the diagrams in figures \ref{fig:3loopdiags}(a) and (b). We remind the reader not to confuse singlet (or non-singlet) diagrams and flavor singlet (or non-singlet) quark currents.  Note that also diagonal flavor non-singlet currents
can lead to non-vanishing singlet diagrams (due to the flavor symmetry breaking effects of the quark masses and condensates).
This can been seen for example from the trace terms $\tr\bigl[\Rra \Rrc\bigr]\tr\bigl[\Rrb \Mm\bigr]$ in $\CO_2^J$ and $\tr\bigl[\Rra \Mmp{2}\bigr]\tr\bigl[\Rrb \Mmp{2}\bigr]$ in $\CO_6^J$.
For two non-singlet currents the related trace coefficients are $c^J_{2,3}$ and $c^J_{6,\{3,4\}}$.
We also note that we find additional finite corrections from singlet diagrams to some of the previously known Wilson coefficients obtained at two loops~\cite{Chetyrkin:1985kn,Surguladze:1990sp}, where contributions from singlet diagrams were omitted, see \subsec{diagonals}.

Interestingly, our results also comprise finite Wilson coefficients for mixed correlators with one flavor singlet and one diagonal non-singlet current or two different diagonal non-singlet currents.
Such transitions between different flavor currents cannot arise in the Wilson coefficient of the identity operator at leading order in the OPE, but they can arise for the higher dimensional OPE
operators which contain non-trivial flavor information. At dimension four these are the quark condensate operators $\op_2^{ij}$ and quark mass operators $\op_6^{ij}$, $\op_6^{ijk}$, as those act as additional flavor sinks and sources.
This is for example the case for the $(\tr\bigl[\Rra \Rrc\bigr]\tr\bigl[\Rrb \Mm\bigr] + \tr\bigl[\Rra \Mm\bigr] \tr\bigl[\Rrb \Rrc\bigr])$ term in $\CO_2^J$ originating from singlet diagrams or
the $(\tr\bigl[\Rra \Mm \Rrb \Rrc\bigr] + \tr\bigl[\Rrb \Mm \Rra \Rrc\bigr])$ term from non-singlet diagrams.
The coefficients contributing to such mixed flavor current correlators are  $c^J_{2,\{3,4,5,6\}}$ and $c^J_{6,\{1,3,4,5,6,7,8,9\}}$. A concrete example is discussed in \subsec{unequal}.
We finally note that the contributions from flavor singlet and non-singlet currents in the $\CO_n^J$ can be separated
by Fierzing all traces involving either $\rho_1$ and/or $\rho_2$ besides other matrices so that $\tr\bigl[\Rra\bigr]$,  $\tr\bigl[\Rrb\bigr]$,  $\tr\bigl[\Rra t^a\bigr]$ and $\tr\bigl[\Rrb t^a\bigr]$ are the only traces containing $\rho_1$ or $\rho_2$.

\section{Discussion of selected results}
\label{sec:discussion}

Let us now discuss our concrete results for a number of relevant cases in QCD with $N_c=3$, $n_f=3$, $T_F=T_F^f=1/2$, $t^a=\lambda^a/2$, where $\lambda^a$ denote the Gell-Mann matrices, and  $\mu^2=-q^2$.
To this end we set $N_A = 8$, $C_F=4/3$, $C_A=3$, $d^{(3)}_{FF} = 5/6$.
In the following, we mostly display rounded numerical results for the sake of brevity, which already allow for interesting observations concerning the convergence of the perturbation series for many low-energy QCD applications. We will briefly comment on a few notable cases.
The complete analytic results for arbitrary renormalization scale $\mu$, gauge group, fermion representation, and total flavor number $n_f$, as well as the results for all cases not elaborated in the following can be obtained from the coefficients listed in \app{ResCoef} or the ancillary file.

\subsection{Flavor off-diagonal currents}

We start with the correlator of the off-diagonal currents
$j_1^J=\bar u \, \Gamma^J d$ and $j_2^J = \bar d \,\Gamma^J u$  (i.e.\ $\rho_{1}=\rho_2^\dagger=t^1 + i t^2$), which for $J=V,A$ is relevant for semileptonic $\tau$ lepton decays. In this case singlet diagrams do not contribute. The results read
\begin{align}
\label{eq:C1vT}
\CO^{V/A,T}_1 = {}&  \biggl\{\! \frac{1}{12} \aop  + \frac{7}{72} \aopn{2} +
\aopn{3} \biggl[ -\frac{1847}{3456}  + \frac{5\zeta_3}{8}  +
n_f \biggl(\! -\frac{59}{1296}
+ \frac{7\zeta_3}{72} \biggr) \biggr]\! \biggr\}\op_1 \notag \\
= {}&  \biggl\{ 0.083 \aop  + 0.097 \aopn{2} + 0.43 \aopn{3}  \biggr\}\op_1 \,, \\
\label{eq:C1vL}
\CO^{V/A,L}_1 = {}& 0\,, \\
\CO^{S/P}_1 = {}& \left\{\! \frac{1}{8} \aop  + \frac{11}{16} \aopn{2}
+ \aopn{3} \left[\frac{13127}{2304}- \frac{17 \zeta_3}{16}
-\frac{229}{864} n_f  \right] \! \right\}\op_1 \notag \\
 = {}& \biggl\{\! 0.13 \aop  + 0.69 \aopn{2} + 3.63 \aopn{3} \! \biggr\}\op_1\,,
\end{align}
for the gluon condensate contributions (keeping $n_f$ explicit) and
\begin{align}
 \CO^{V/A,T}_2 = {}& - \left[ \aop + 3.73 \aopn{2} + 17.29 \aopn{3} \right] (m_u \bar{u}u + m_d \bar{d}d)\notag\\
 & + \left[ 0.60 \aopn{2} + 6.58 \aopn{3} \right] m_s \bar{s}s  \notag\\
 & \pm  \left[ 1+ 1.33 \aop + 9.83 \aopn{2} + 82.12  \aopn{3} \right] (m_d \bar{u}u + m_u \bar{d}d) \,,\\
 \CO^{V/A,L}_2 = {}& (m_u \mp m_d) \bar{u}u + (m_d \mp m_u) \bar{d}d \,, \\
 \CO^{S/P}_2 = {}&  + \left[0.5 + 1.83 \aop + 11.29 \aopn{2} + 73.13	 \aopn{3} \right] (m_u \bar{u}u + m_d \bar{d}d) \notag\\
 & -  \left[ 0.83 \aopn{2} + 13.64 \aopn{3} \right] m_s \bar{s}s  \notag\\
 & \pm  \left[ 1+ 4.67 \aop + 38.28 \aopn{2} + 355.14  \aopn{3} \right] (m_d \bar{u}u + m_u \bar{d}d) \,,
\\[2 ex]
  \CO^{V/A,T}_6 = {}
 & -\left[ 0.038 + 0.076 \aop + 0.46 \aopn{2} \right] \left(m_u^4 + m_d^4\right)
   -0.047 \aopn{2} m_s^4 \notag\\
 & +\left[0.077 \aop + 1.05 \aopn{2}\right] m_u^2 m_d^2
   + 0.089 \aopn{2} \left(m_u^2 m_s^2 + m_d^2 m_s^2\right) \notag\\
 & \pm\left[0.011 \aop + 0.53 \aopn{2}\right] \left(m_u^3 m_d + m_u m_d^3 \right)
   \pm 0.11 \aopn{2} m_u m_d m_s^2 \,,\\
 \CO^{V/A,L}_6 = {}
& - \left[ 0.076 + 0.27 \aop + 2.13 \aopn{2} \right] \left(m_u^4+m_d^4\right) \notag\\
& + \left[ 0.15 + 0.96 \aop + 9.72 \aopn{2}   \right] m_u^2 m_d^2
  + 0.068 \aopn{2} \left(m_u^2 m_s^2+m_d^2 m_s^2\right) \notag\\
& \mp \left[ 0.21 \aop + 2.72 \aopn{2} \right] \left(m_u^3 m_d+m_u m_d^3\right)
  \mp 0.14 \aopn{2} m_u m_d m_s^2 \,,\\
 \CO^{S/P}_6 = {}
& + \left[ 0.019 + 0.069 \aop + 0.27 \aopn{2} \right] \left(m_u^4+m_d^4\right)
- 0.0036 \aopn{2} m_s^4 \notag\\
& + \left[ 0.076 + 0.57 \aop  + 4.78 \aopn{2} \right] m_u^2 m_d^2
- 0.36 \aopn{2} \left(m_u^2 m_s^2+m_d^2 m_s^2\right) \notag\\
& \pm \left[ 0.061 \aop + 1.11 \aopn{2} \right] \left(m_u^3 m_d+m_u m_d^3\right)
\mp 0.43 \aopn{2} m_u m_d m_s^2 \,,
\end{align}
for the quark condensate and the mass corrections.
All results are in full agreement with \rcite{Chetyrkin:1985kn} to two loops and \rcites{HarlanderDiss,Harlander:2020duo} to three loops.

\subsection{Flavor diagonal $\rho$ meson currents}

For the two diagonal vector currents $j_1^V=j_2^V=\frac{1}{\sqrt{2}}(\bar u\gamma^\mu u-\bar d\gamma^\mu d)$ (i.e.\ $\rho_{1}=\rho_2=\sqrt{2\,}t^3$) creating/annihilating a  $\rho$ meson, the gluon condensate Wilson coefficients agree with \eqs{C1vT}{C1vL} and the quark condensate and quark mass terms read
\begin{align}
\CO^{V,T}_2 ={}
& + \left[ 1 + 0.33 \aop + 6.10 \aopn{2} + 64.83 \aopn{3} \right](m_u \bar u u + m_d \bar d d) \notag\\
& +\left[  0.60 \aopn{2} + 6.58 \aopn{3} \right]\,m_s \bar ss  \,,\\
\CO^{V,T}_6   ={}
 & + \left[ - 0.038 - 0.027 \aop + 0.45 \aopn{2}   \right] \left(m_u^4+m_d^4\right)
   - 0.047 \aopn{2} m_s^4 \notag\\
 & + 0.29 \aopn{2} m_u^2 m_d^2
   + 0.14 \aopn{2} \left(m_u^2 m_s^2+m_d^2 m_s^2\right) \,,\\
\CO^{V,L}_2={}& \CO^{V,L}_6=0 \,.
\end{align}
The quark condensate terms to $\ord{\alpha_s^2}$ agree with \rcite{Chetyrkin:1985kn}. The corresponding $\ord{\alpha_s^3}$ terms and all the quark mass terms are to the best of our knowledge new.
There are no contributions from singlet diagrams.

\subsection{Flavor diagonal strange quark currents}
\label{subsec:diagonals}

The correlator of the two vector currents $j_1^V=j_2^V=\bar s\gamma^\mu s$  (i.e.\ $\rho_{1}=\rho_2=\frac{1}{3}\bbid -\frac{2}{\sqrt{3}}t^8$)
also receives contributions from singlet diagrams.
However, they only arise at the three-loop level and do not contribute to the previously known two-loop results. For the gluon condensate Wilson coefficients we obtain
\begin{align}
\label{eq:C1vTss}
\CO^{V,T}_1 = {}&  \left\{ \frac{1}{12} \aop + \frac{7}{72} \aopn{2} +
\aopn{3} \left[-\frac{1627}{3456}  + \frac{35\zeta_3}{72}  +
n_f \left(-\frac{59}{1296}
+ \frac{7\zeta_3}{72} \right) \right] \right\}\,\op_1  \notag \\
 = {}& \left\{ 0.083 \aop + 0.097 \aopn{2} + 0.33\aopn{3}  \right\}\,\op_1  \,, \\
\CO^{V,L}_1 ={}& 0 \,.
\end{align}
Here, the result for the transverse Wilson coefficient differs from \eq{C1vT} at $\ord{\alpha_s^3}$ due to contributions from three-loop singlet diagrams contained in the trace coefficient $c_{1,1}^{V,T}$.
If the singlet diagrams are omitted the results agree with \eq{C1vT}.
The quark condensate results read
\begin{align}
\label{eq:C2vTss}
\CO^{V,T}_2 = {}
 & +\left[0.60 \aopn{2} + 6.58 \aopn{3} \right](m_u \bar{u} u + m_d \bar{d} d) \notag\\
 & + \left[ 2 + 0.67 \aop + 11.60 \aopn{2} +  117.14 \aopn{3} \right] \,m_s \bar{s}s \,,
\\
\CO^{V,L}_2 = {}& 0 \,.
\end{align}
In $\CO^{V,T}_2$ a three-loop singlet diagram correction contained in the trace coefficient $c_{2,4}^{V,T}$ contributes to the $\ord{\alpha_s^3}$ coefficient of the $m_s\bar ss$ condensate, which would read $123.08$ in stead of $117.14$, if the singlet diagrams are omitted.
Up to $\ord{\alpha_s^2}$ singlet diagrams do not contribute and the results up to this order agree with the previously known $\ord{\alpha_s^2}$ results from \rcites{Chetyrkin:1985kn,Surguladze:1990sp}.
For the quark mass terms no singlet diagrams contribute up to and including
 $\ord{\alpha_s^2}$. The results read
 \begin{align}
\label{eq:C6vTss}
\CO^{V,T}_6 = {}
 & + \left[ - 0.076 - 0.053 \aop  +0.94 \aopn{2} \right] m_s^4
   - 0.047 \aopn{2} \left(m_u^4+m_d^4\right) \notag\\
 & + 0.29 \aopn{2} \left(m_u^2 m_s^2+m_d^2 m_s^2\right) \,,\\
\CO^{V,L}_6 ={}& 0\,.
\end{align}
The $\ord{\alpha_s}$ contributions have been determined before in \rcite{Chetyrkin:1985kn}, and we agree. We note that the series for the transversal $m_s^4$ contribution does not exhibit good convergence for $\as(m_\tau)/\pi \approx 0.075$.

\bigskip

For two  axial-vector currents $j_1^A=j_2^A= \bar s\gamma^\mu\gamma_5 s$
singlet diagrams contribute already
at two loops, i.e.\ one loop order lower than for the vector current case.
These singlet diagram contributions were not included in the results of \rcite{Chetyrkin:1985kn}. For the gluon condensate Wilson coefficients we obtain
\begin{align}
\label{eq:C1aTss}
\CO^{A,T}_1 = {}&
\left\{ \frac{1}{12} \aop + \frac{5}{36} \aopn{2} +
\aopn{3} \left [-\frac{1331}{3456}  + \frac{23 \zeta_3}{24} + n_f \left(-\frac{37}{648} + \frac{7 \zeta_3}{72} \right) \right]
\right\}\, \op_1 \notag  \\
= {}& \left\{  0.083 \aop + 0.14 \aopn{2} +  0.95 \aopn{3}  \right\}\, \op_1\,,  \\
\label{eq:C1aLss}
\CO^{A,L}_1 = {}& \left\{
-\frac{1}{4} \aopn{2} + \aopn{3} \left[ -\frac{157}{48} + \frac{5 n_f}{72} \right]
\right\} \, \op_1 \notag \\
= {}& \left\{ - 0.25 \aopn{2} - 3.06 \aopn{3} \right\} \, \op_1 \,.
\end{align}
The results agree with \rcite{Chetyrkin:1985kn} at the $\ord{\alpha_s}$ level, but already differ at $\ord{\alpha_s^2}$ as anticipated.
The non-zero longitudinal Wilson coefficient in \eq{C1aLss} arises entirely from singlet diagrams. It is interesting to observe that for $\as(m_\tau)/\pi \approx 0.075$ its convergence is only marginal in contrast to the good convergence of all other gluon condensate Wilson coefficients displayed here.
The results for the quark condensate and quark mass contributions read
\begin{align}
\label{eq:C2aTss}
\CO^{A,T}_2 = {}
 & - \left[2 +  4.67 \aop + 22.75 \aopn{2} + 147.70 \aopn{3} \right] \,m_s \bar ss \notag\\
 & + \left[0.60 \aopn{2} + 5.89 \aopn{3} \right] (m_u \bar u u + m_d \bar d d) \,,\\
\label{eq:C2aLss}
\CO^{A,L}_2 = {}
 & +  \left[ 4 - 12 \aopn{2} - 112.71 \aopn{3} \right] \,m_s \bar ss - 0.33 \aopn{3} \, (m_u \bar u u + m_d \bar d d)  \,,\\
\label{eq:C6atss}
\CO^{A,T}_6 = {}
 & - \left[ 0.076  + 0.096 \aop + 0.17 \aopn{2} \right] m_s^4
   - 0.047 \aopn{2} \left(m_u^4+m_d^4\right) \notag\\
 & + 0.070 \aopn{2} \left(m_u^2 m_s^2+m_d^2 m_s^2\right) \,,\\
\label{eq:C6aLss}
\CO^{A,L}_6 = {}
& + \left[ 0.85 \aop  + 8.91 \aopn{2} \right] m_s^4 + 0.27 \aopn{2} \left(m_u^2 m_s^2+m_d^2 m_s^2\right) \,.
\end{align}
For $\CO^{A,T}_2$, like for the gluon condensate results, we differ from the result given in \rcite{Chetyrkin:1985kn} already at $\ord{\alpha_s^2}$ due to singlet diagrams. For $\CO^{A,T}_6$ and $\CO^{A,L}_6$ there is agreement up to and including $\ord{\alpha_s}$ since here singlet diagrams do not yet contribute at the two-loop level.
No result was quoted in \rcite{Chetyrkin:1985kn} for $\CO^{A,L}_2$.
Note also that for $\CO^S_2$ we agree with \rcite{Surguladze:1990sp} to $\ord{\as^2}$ again only when omitting singlet diagrams.

\subsection{Unequal flavor diagonal currents}
\label{subsec:unequal}

Finally, let us discuss the case of a correlator with two different flavor diagonal currents. As an example we consider
$j^J_1=\frac{1}{\sqrt{2}}(\bar u \Gamma^J u-\bar d \Gamma d)$  and $j^J_2=\frac{1}{\sqrt{6}}(\bar u \Gamma^J u + \bar d \Gamma^J d - 2\bar s \Gamma^J s)$ (i.e.\ $\rho_1=\sqrt{2}\,t^3$, $\rho_2=\sqrt{2}\,t^8$). All gluon condensate Wilson coefficients (like the Wilson coefficients of the leading identity operator) vanish due to flavor symmetry in the massless quark limit. The quark condensate und quark mass Wilson coefficients, on the other hand, are non-zero due to the breaking effects of the finite quark masses and quark condensates (or equivalently the flavor content of the matrices $M$ and $Q$). Here, both singlet as well as non-singlet diagrams contribute and the results read
\begin{align}
\CO^{V,T}_2 = {}
& +\left[ 0.58 + 0.19 \aop + 3.18 \aopn{2} + 33.63 \aopn{3} \right] \,(m_u \bar u u - m_d \bar d d) \,,\\
\CO^{V,L}_2 = {}& 0 \,,\\
\CO^{A,T}_2 = {}&  - \left[0.58 + 1.35 \aop + 8.18 \aopn{2} + 61.19 \aopn{3} \right] \,(m_u \bar u u - m_d \bar d d) \,,\\
\CO^{A,L}_2 = {}& 1.15 \,  (m_u \bar u u - m_d \bar d d) \,,\\
\CO^{S}_2 = {}
 & + \left[ 0.87 + 3.75 \aop + 26.67 \aopn{2} + 205.45 \aopn{3} \right]\,(m_u \bar u u - m_d \bar d d) \nonumber\\
 & + \left[ 2.43 \aopn{2} + 49.69 \aopn{3} \right] \, ( m_u \bar s s - m_d \bar s s + m_s \bar u u - m_s \bar d d ) \,,\\
\CO^{P}_2 = {}
 &  - \left[ 0.29 + 1.64 \aop + 10.36 \aopn{2} + 70.93 \aopn{3} \right]\,(m_u \bar u u - m_d \bar d d) \nonumber\\
 & - \left[ 4.74 \aopn{2} + 84.01 \aopn{3} \right] \, ( m_u \bar s s - m_d \bar s s + m_s \bar u u - m_s \bar d d )\,,
\end{align}
and
\begin{align}
\CO^{V,T}_6 = {}
 & \left[ -0.022 - 0.015 \aop + 0.29 \aopn{2} \right] (m_u^4-m_d^4) + 0.083 \aopn{2}   (m_u^2-m_d^2) m_s^2\,,\\
\CO^{V,L}_6 = {}& 0\,,\\
\CO^{A,T}_6 = {}
 & \left[ -0.022 - 0.028 \aop + 0.027 \aopn{2} \right] (m_u^4-m_d^4) -0.55 \aopn{2} m_s^2 (m_u^2-m_d^2) \,,\\
\CO^{A,L}_6 = {}
 & \left[ 0.25 \aop  +  2.48 \aopn{2}  \right] (m_u^4-m_d^4) + 1.25 \aopn{2}   (m_u^2-m_d^2) m_s^2 \,,\\
\CO^{S}_6 = {}
 &  \left[ 0.033 + 0.24 \aop + 2.37 \aopn{2} \right]\left( m_u^4 - m_d^4\right)
 - 0.34 \aopn{2} \left( m_u^2 - m_d^2 \right) m_s^2 \notag\\
 & + 0.15 \aopn{2} \left( m_u^3 m_s + m_u m_s^3 - m_d^3 m_s - m_d m_s^3\right) \,,\\
\CO^{P}_6 = {}
 &  \left[ 0.033 + 0.17 \aop + 0.68 \aopn{2} \right]\left( m_u^4 - m_d^4\right)
 - 0.085 \aopn{2} \left( m_u^2 - m_d^2 \right) m_s^2 \notag\\
 & + 0.30 \aopn{2} \left( m_u^3 m_s + m_u m_s^3 - m_d^3 m_s - m_d m_s^3\right) \,.
\end{align}

\section{Conclusions and outlook}
\label{sec:conclusions}

In this article we have calculated the three-loop Wilson coefficients of the physically relevant dimension-four operators $G_{\mu\nu}^a G^{a,\mu\nu}$, $m_i\bar q_j q_j$ and $m_i m_j m_k^2$ in the short-distance operator product expansion of the time-ordered product of two gauge-singlet vector, axial-vector, scalar, and pseudo-scalar currents. The results are provided for an arbitrary light quark flavor content of each of the two currents and for a general non-Abelian gauge group with the fermions (quarks) living in an arbitrary representation.
We have fully accounted for the contributions from flavor singlet currents and from so-called singlet diagrams.
To this end we have employed the Larin scheme for the treatment of $\gamma_5$ in the definition of axial-vector and pseudo-scalar currents, which is consistent within dimensional regularization and includes additional finite renormalization factors to restore the chiral Ward identities.
We have given the results in a compact notation based on current flavor density matrices in \eqsm{CO1}{CO6} with the corresponding trace coefficients listed in \app{ResCoef}. This allows to also determine the non-vanishing Wilson coefficients for currents with different flavor content arising from the flavor symmetry breaking effects of the quark condensates and masses. All results are also provided in an ancillary file in \texttt{Mathematica} readable format.

With our calculational setup based on the \texttt{Looping} code and the available multi-loop technology, see e.g.\ \rcite{Ruijl:2017cxj}, we believe that also the four-loop results for the Wilson coefficients addressed in this paper are within reach.
This computation is left to upcoming work.

\begin{acknowledgments}
We acknowledge the kind support of the  Erwin-Schr\"odinger Institute for Mathematics and Physics during the thematic program "Quantum Field Theory at the Frontiers of the Strong Interaction" taking place July 31 - September 1, 2023.
RB and MS would like to thank the Particle Physics Group of the University of Vienna for hospitality, and AHH would like to thank the Particle Physics Group at the Albert-Ludwigs-Universit\"at Freiburg for hospitality.
RB was supported by the German Research Foundation (DFG) under grant DI 784/6-1. 
We acknowledge partial support by the FWF Austrian Science Fund under the Project No.~P32383-N27.
The Feynman graphs in this paper were drawn using \texttt{JaxoDraw}~\cite{Binosi:2008ig} and \texttt{TikZ}~\cite{Tan12}.
\end{acknowledgments}

\begin{appendix}

\section{Renormalization constants}
\label{app:renconst}

{\scriptsize
\begin{flalign}
  Z_\alpha ={}&
  1
  +\frac{\as}{4\pi} \biggl(
  -\frac{11 C_A}{3 \epsilon }
  +\frac{4 n_f T_F}{3 \epsilon }\biggr)
  + \biggl(\frac{\as}{4\pi} \biggr)^{\!2}
  \biggl(\frac{121 C_A^2}{9 \epsilon ^2}
  -\frac{88 C_A n_f T_F}{9 \epsilon ^2}
  +\frac{16 n_f^2 T_F^2}{9 \epsilon ^2}
  -\frac{17 C_A^2}{3 \epsilon }
  +\frac{10 C_A n_f T_F}{3 \epsilon}
  \nn\\
  &\quad
  +\frac{2 C_F n_f T_F}{\epsilon }\biggr)
  + \ord{\as^3} \,,
  \\[1 ex]
  Z_m ={}&
  1 - \frac{\as}{4\pi} \frac{3 C_F}{\epsilon }
  +\biggl(\frac{\as}{4\pi} \biggr)^{\!2} \biggl(
  \frac{11 C_A C_F}{2 \epsilon ^2}+\frac{9 C_F^2}{2 \epsilon ^2}
  -\frac{2 C_F n_f T_F}{\epsilon ^2}
  -\frac{97 C_A C_F}{12 \epsilon }-\frac{3 C_F^2}{4 \epsilon }
  +\frac{5 C_F n_f T_F}{3 \epsilon }\biggr)
  \nn\\
  &
  +\biggl(\frac{\as}{4\pi} \biggr)^{\!3}\biggl(
  -\frac{121 C_A^2 C_F}{9 \epsilon ^3}
  -\frac{33 C_A C_F^2}{2 \epsilon ^3}
  -\frac{9 C_F^3}{2 \epsilon ^3}
  +\frac{88 C_A C_F n_f T_F}{9 \epsilon ^3}
  +\frac{6 C_F^2 n_f T_F}{\epsilon ^3}
  -\frac{16 C_F n_f^2 T_F^2}{9 \epsilon ^3}
  +\frac{1679 C_A^2 C_F}{54 \epsilon ^2}
  \nn\\
  &\quad
  +\frac{313 C_A C_F^2}{12 \epsilon ^2}
  +\frac{9 C_F^3}{4 \epsilon ^2}
  -\frac{484 C_A C_F n_f T_F}{27 \epsilon ^2}
  -\frac{29 C_F^2 n_f T_F}{3 \epsilon ^2}
  +\frac{40 C_F n_f^2 T_F^2}{27 \epsilon ^2}
  -\frac{11413 C_A^2 C_F}{324 \epsilon }
  +\frac{43 C_A C_F^2}{4 \epsilon }
  \nn\\
  &\quad
  -\frac{43 C_F^3}{2 \epsilon }
  +\frac{556 C_A C_F n_f T_F}{81 \epsilon }
  +\frac{46 C_F^2 n_f T_F}{3 \epsilon }
  +\frac{140 C_F n_f^2 T_F^2}{81 \epsilon }
  +\frac{16 C_A C_F n_f T_F \zeta_3}{\epsilon }
  -\frac{16 C_F^2 n_f T_F \zeta_3}{\epsilon }
  \biggr)
  \nn\\
  &
  + \ord{\as^4} \,,
  \\[1 ex]
  Z_{11} ={}&
  1
  +\frac{\as}{4\pi} \biggl(-\frac{11 C_A}{3 \epsilon }+\frac{4 n_f T_F}{3 \epsilon }\biggr)
  + \biggl(\frac{\as}{4\pi} \biggr)^{\!2}\biggl(
  \frac{121 C_A^2}{9 \epsilon ^2}
  -\frac{88 C_A n_f T_F}{9 \epsilon ^2}
  +\frac{16 n_f^2 T_F^2}{9 \epsilon ^2}
  -\frac{34 C_A^2}{3 \epsilon }
  +\frac{20 C_A n_f T_F}{3 \epsilon}
  \nn\\
  &\quad
  +\frac{4 C_F n_f T_F}{\epsilon }
  \biggr)
  + \ord{\as^3} \,,
  \\[1 ex]
  Z_{12} ={}&
  \frac{\as}{4\pi} \frac{12 C_F}{\epsilon }
  +\biggl(\frac{\as}{4\pi} \biggr)^{\!2}  \biggl(-\frac{44 C_A C_F}{\epsilon ^2}+\frac{16 C_F n_f T_F}{\epsilon ^2}+\frac{194 C_A C_F}{3 \epsilon }+\frac{6
    C_F^2}{\epsilon }-\frac{40 C_F n_f T_F}{3 \epsilon }\biggr)
  + \ord{\as^3} \,,
  \\[1 ex]
  Z_0^{\rd} ={}&\frac{N_c}{16 \pi^2} \Biggl\{
  \frac{1}{\epsilon }
  + \frac{\as}{4\pi} \biggl(-\frac{6 C_F}{\epsilon ^2}+\frac{2 C_F}{\epsilon }\biggr)
  + \biggl(\frac{\as}{4\pi} \biggr)^{\!2} \biggl(\frac{22 C_A C_F}{3 \epsilon ^3}
  +\frac{24 C_F^2}{\epsilon ^3}
  -\frac{8 C_F n_f T_F}{3 \epsilon ^3}
  -\frac{24 C_A C_F}{\epsilon ^2}
  -\frac{10 C_F^2}{\epsilon ^2}
    \nn\\
  &\quad
  +\frac{16  C_F n_f T_F}{3 \epsilon ^2}
  +\frac{109 C_A C_F}{6 \epsilon }
  -\frac{131 C_F^2}{6 \epsilon }
  -\frac{16 C_F T_F}{\epsilon}
   -\frac{10 C_F n_f T_F}{3 \epsilon }
  -\frac{8 C_A C_F \zeta_3}{\epsilon }
  +\frac{16 C_F^2 \zeta_3}{\epsilon }\biggr)
  + \ord{\as^3}
  \Biggr\}\,,
  \\[1 ex]
  Z_0^{\rnd} ={}&
  \frac{N_c}{16 \pi^2}  \biggl(\frac{\as}{4\pi} \biggr)^{\!2}
   \biggl(- \frac{16 C_F T_F}{\eps} \biggr)+ \ord{\as^3} \,,
  \\[1 ex]
  Z_\msb^\rs   ={}&
  1+  \biggl(\frac{\as}{4\pi} \biggr)^{\!2} \biggl(
  \frac{22C_AC_F}{3 \epsilon }   +\frac{10C_Fn_fT_F}{3 \epsilon }\biggr)
  + \biggl(\frac{\as}{4\pi} \biggr)^{\!3} \biggl(
  -\frac{484C_A^2C_F}{27 \epsilon ^2}
  -\frac{44C_AC_Fn_fT_F}{27 \epsilon ^2}
  +\frac{80C_Fn_f^2T_F^2}{27 \epsilon ^2}
    \nn\\
  &\quad
  +\frac{3578C_A^2 C_F}{81 \epsilon }
  -\frac{308C_AC_F^2}{9 \epsilon }
  +\frac{298C_AC_Fn_fT_F}{81 \epsilon }
  -\frac{44C_F^2n_f T_F}{9 \epsilon }
  +\frac{104C_Fn_f^2T_F^2}{81 \epsilon }
  \biggr)
  +\ord{\as^4} \,,
  \\[1 ex]
  Z_5^\rs  ={}&
  1- \frac{\as}{4\pi}  4C_F
  +\biggl(\frac{\as}{4\pi} \biggr)^{\!2} \biggl(\!
  -\frac{107}{9}C_AC_F
  +22C_F^2
  +\frac{31}{9}C_Fn_fT_F \biggr)
  +\biggl(\frac{\as}{4\pi} \biggr)^{\!3} \biggl(\!
  -\frac{2147}{27}C_A^2C_F
  +\frac{5834}{27}C_AC_F^2
      \nn\\
  &\quad
  -\frac{370}{3}C_F^3
  -\frac{266}{81}C_AC_Fn_fT_F
  +\frac{497}{27}C_F^2  n_f T_F
  +\frac{1264}{81}C_Fn_f^2T_F^2
  +56C_A^2C_F \zeta_3
  -160C_AC_F^2 \zeta_3
  +96C_F^3  \zeta_3
  \nn\\
  &\quad
  +\frac{220}{3}C_A C_F n_f T_F \zeta_3
  -\frac{208}{3}C_F^2n_fT_F \zeta_3
  \biggr)
  +\ord{\as^4} \,,
  \\[1 ex]
  Z_\msb^\rns  ={}&
  1+\biggl(\frac{\as}{4\pi} \biggr)^{\!2}  \biggl(
  \frac{22 C_A C_F}{3 \epsilon }
  -\frac{8 C_F n_f T_F}{3 \epsilon }\biggr)
  +\biggl(\frac{\as}{4\pi} \biggr)^{\!3}  \biggl(
  -\frac{484 C_A^2 C_F}{27 \epsilon^2}
  +\frac{352 C_A C_F n_f T_F}{27 \epsilon ^2}
  -\frac{64 C_F n_f^2 T_F^2}{27 \epsilon ^2}
  \nn\\
  &\quad
  +\frac{3578 C_A^2  C_F}{81 \epsilon }
  -\frac{308 C_A C_F^2}{9 \epsilon }
  -\frac{1664 C_A C_F n_f T_F}{81 \epsilon }
  +\frac{64 C_F^2 n_f T_F}{9 \epsilon }
  +\frac{32 C_F n_f^2 T_F^2}{81 \epsilon }
  \biggr)
  +\ord{\as^4} \,,
  \\[1 ex]
  Z_5^\rns  ={}&
  1-   \frac{\as}{4\pi} 4C_F
  +\biggl(\frac{\as}{4\pi} \biggr)^{\!2}  \biggl(-\frac{107}{9}C_AC_F +22C_F^2+\frac{4}{9}C_F n_fT_F \biggr)
  +\biggl(\frac{\as}{4\pi} \biggr)^{\!3}  \biggl(
  -\frac{2147}{27} C_A^2 C_F
  +\frac{5834}{27} C_AC_F^2
  \nn\\
  &\quad
  -\frac{370}{3} C_F^3
  +\frac{712}{81}C_AC_F n_fT_F
  -\frac{124}{27}C_F^2  n_fT_F
  +\frac{208}{81}C_F n_f^2T_F^2
  +56C_A^2C_F \zeta_3
  -160C_AC_F^2 \zeta_3
  +96C_F^3\zeta_3
  \nn\\
  &\quad
  +\frac{64}{3}C_AC_F n_fT_F \zeta_3
  -\frac{64}{3}C_F^2 n_fT_F \zeta_3\biggr)
  +\ord{\as^4} \,,
  \\[1 ex]
  Z_\msb^P  ={}&
  1- \frac{\as}{4\pi} \frac{3 C_F}{\epsilon }
  +\biggl(\frac{\as}{4\pi} \biggr)^{\!2}  \biggl(
  \frac{11 C_A C_F}{2 \epsilon ^2}
  +\frac{9 C_F^2}{2 \epsilon ^2}
  -\frac{2 C_F n_f T_F}{\epsilon ^2}
  +\frac{79 C_A C_F}{12 \epsilon }
  -\frac{3 C_F^2}{4 \epsilon }
  -\frac{11 C_F n_f T_F}{3 \epsilon }\biggr)
  \nn\\
  &
  +\biggl(\frac{\as}{4\pi} \biggr)^{\!3}
  \biggl(-\frac{121 C_A^2 C_F}{9 \epsilon ^3}
  -\frac{33 C_A C_F^2}{2 \epsilon ^3}
  -\frac{9 C_F^3}{2 \epsilon ^3}
  +\frac{88 C_A C_F n_f T_F}{9 \epsilon ^3}
  +\frac{6 C_F^2 n_f T_F}{\epsilon ^3}
  -\frac{16 C_F n_f^2 T_F^2}{9 \epsilon ^3}
  -\frac{257 C_A^2 C_F}{54 \epsilon ^2}
  \nn\\
  &\quad
  -\frac{215 C_A C_F^2}{12 \epsilon ^2}
  +\frac{9 C_F^3}{4 \epsilon ^2}
  +\frac{220 C_A C_F n_f T_F}{27 \epsilon ^2}
  +\frac{19 C_F^2 n_f T_F}{3 \epsilon ^2}
  -\frac{88 C_F n_f^2 T_F^2}{27 \epsilon ^2}
  -\frac{599 C_A^2 C_F}{108 \epsilon }
  +\frac{3203 C_A C_F^2}{36 \epsilon }
  \nn\\
  &\quad
  -\frac{43 C_F^3}{2 \epsilon }
  -\frac{116 C_A C_F n_f T_F}{9 \epsilon }
  -\frac{214 C_F^2 n_f T_F}{9 \epsilon }
  +\frac{68 C_F n_f^2 T_F^2}{27 \epsilon }
  +\frac{16 C_A C_F n_f  T_F \zeta_3}{\epsilon }
  -\frac{16 C_F^2 n_f T_F \zeta_3}{\epsilon }
  \biggr)
  \nn\\
  &
  + \ord{\as^4} \,,
  \\[1 ex]
  Z_5^P  ={}&
  1- \frac{\as}{4\pi} 8 C_F
  +\biggl(\frac{\as}{4\pi} \biggr)^{\!2} \biggl(
  \frac{2}{9} C_A C_F
  +\frac{8}{9} C_F n_f T_F \biggr)
  +\biggl(\frac{\as}{4\pi} \biggr)^{\!3}  \biggl(
  -\frac{958}{27} C_A^2 C_F
  -\frac{800}{27} C_A C_F^2
  +\frac{304}{3} C_F^3
  \nn\\
  &\quad
  +\frac{1712}{81} C_A C_F n_f T_F
  -\frac{1160}{27} C_F^2 n_f T_F
  +\frac{416}{81} C_F n_f^2 T_F^2
  -208 C_A^2 C_F \zeta_3
  +608 C_A C_F^2 \zeta_3
  -384 C_F^3 \zeta_3
  \nn\\
  &\quad
  +\frac{128}{3} C_A C_F n_f T_F \zeta_3
  -\frac{128}{3} C_F^2 n_f T_F \zeta_3
  \biggr)
  + \ord{\as^4} \,.
\end{flalign}
}


\section{Results for the Wilson coefficients}
\label{app:ResCoef}

In this appendix we present our analytic results of the coefficients in \eqsm{CO1}{CO6}.
As mentioned above, we adopt a notation, where the gauge group is SU($N_c$) and the fermions are in the fundamental representation $F$.
In order to manifestly generalize our results to an arbitrary (compact semi-simple) gauge group and an arbitrary irreducible fermion representation $R$ with dimension $N_R$ one needs to replace
$T_F \to T_R$, $N_c \to N_R$, $C_F \to C_R$, and $d^{(3)}_{FF} \to d^{(3)}_{RR}$ in the $c_{n,m}$ coefficients and $N_c \to N_R$ in \eq{CO6}.
Furthermore we define
\begin{align}
  L_q \equiv  \ln \biggl( \frac{\mu^2}{-q^2} \biggr) \,.
\end{align}
For convenience of the reader we provide our results also in a \texttt{Mathematica} readable file supplementing this publication.

\subsection{Gluon condensate}

\subsubsection{Vector and axial-vector currents}

{\scriptsize
\input{Coefficients_C1-VA_wot.tex}
}

\subsubsection{Scalar and pseudo-scalar currents}

{\scriptsize
\input{Coefficients_C1-SP_wot.tex}
}

\subsection{Quark condensates}

\subsubsection{Vector and axial-vector currents}

{\scriptsize
\input{Coefficients_CO2-VA_wot.tex}
}

\subsubsection{Scalar and pseudo-scalar currents}

{\scriptsize
\input{Coefficients_CO2-SP_wot.tex}
}

\subsection{Mass corrections}

\subsubsection{Vector and axial-vector currents}

{\scriptsize
\input{Coefficients_CO6-VA_wot.tex}
}

\subsubsection{Scalar and pseudo-scalar currents}

{\scriptsize
\input{Coefficients_CO6-SP_wot.tex}
}

\end{appendix}

\bibliography{./sources}
\bibliographystyle{JHEP}

\end{document}

%% file: Coefficients_C1-VA_wot.tex

\begin{flalign}
%
&c_{1,1}^{V,L} = c_{1,2}^{V,L} = c_{1,2}^{A,L} = 0\\[3ex] 
%
%
%
& c_{1,1}^{V,T} = \left(\frac{\alpha_s}{\pi}\right)^{\!3}\biggl[\biggl(\frac{11}{18}
-\frac{4 \zeta_3}{3}\biggr) \frac{\dFFF}{N_A}\biggr]&\\[3ex] 
& c_{1,1}^{A,T} = \left(\frac{\alpha_s}{\pi}\right)^{\!2}\biggl[\frac{1}{6} T_F^2\biggr]
+\left(\frac{\alpha_s}{\pi}\right)^{\!3}\biggl[\biggl(\frac{11 L_q}{72}
+\frac{139}{216}\biggr) C_A T_F^2
+\biggl(\zeta_3
-\frac{L_q}{4}
-1\biggr) C_F T_F^2
+\biggl(-\frac{L_q}{18}
-\frac{5}{54}\biggr) n_f T_F^3\biggr]&\\[3ex] 
& c_{1,1}^{A,L} = \left(\frac{\alpha_s}{\pi}\right)^{\!2}\biggl[-T_F^2\biggr]
+\left(\frac{\alpha_s}{\pi}\right)^{\!3}\biggl[\biggl(-\frac{11 L_q}{12}
-\frac{157}{36}\biggr) C_A T_F^2
+\biggl(\frac{L_q}{3}
+\frac{5}{9}\biggr) n_f T_F^3\biggr]&\\[3ex] 
& c_{1,2}^{V,T} = c_{1,2}^{A,T} = \frac{\alpha_s}{\pi}\biggl[\frac{1}{6} T_F\biggr]
+\left(\frac{\alpha_s}{\pi}\right)^{\!2}\biggl[\frac{1}{12} C_A T_F
-\frac{1}{24} C_F T_F\biggr]
+\left(\frac{\alpha_s}{\pi}\right)^{\!3}\biggl[\biggl(\frac{\zeta_3}{18}
+\frac{L_q}{24}
-\frac{1}{72}\biggr) C_A n_f T_F^2
+\biggl(\frac{5 \zeta_3}{36}
-\frac{L_q}{24}
&\notag\\&\quad
-\frac{197}{576}\biggr) C_A^2 T_F
+\biggl(\frac{143}{216}
-\frac{11 L_q}{288}\biggr) C_A C_F T_F
+\biggl(\frac{\zeta_3}{6}
+\frac{L_q}{18}
-\frac{91}{864}\biggr) C_F n_f T_F^2
-\frac{23}{64} C_F^2 T_F\biggr]& 
\end{flalign}

%% file: Coefficients_C1-SP_wot.tex

\begin{flalign}
%
&c_{1,1}^{S} = c_{1,1}^{P} = 0\\[3ex] 
%
%
%
& c_{1,2}^{S} = c_{1,2}^{P} = \frac{\alpha_s}{\pi}\biggl[\frac{1}{4} T_F\biggr]
+\left(\frac{\alpha_s}{\pi}\right)^{\!2}\biggl[\frac{3}{8} C_A T_F
+\biggl(\frac{3 L_q}{8}
+\frac{3}{16}\biggr) C_F T_F\biggr]
+\left(\frac{\alpha_s}{\pi}\right)^{\!3}\biggl[\biggl(-\frac{L_q}{48}
-\frac{11}{72}\biggr) C_A n_f T_F^2
&\notag\\&\quad
+\biggl(-\frac{9 \zeta_3}{32}
+\frac{11 L_q^2}{64}
+\frac{119 L_q}{96}
+\frac{1729}{2304}\biggr) C_A C_F T_F
+\biggl(\frac{L_q}{6}
+\frac{901}{1152}\biggr) C_A^2 T_F
+\biggl(-\frac{L_q^2}{16}
-\frac{5 L_q}{48}
-\frac{65}{144}\biggr) C_F n_f T_F^2
&\notag\\&\quad
+\biggl(-\frac{9 \zeta_3}{16}
+\frac{9 L_q^2}{32}
+\frac{21 L_q}{64}
+\frac{195}{256}\biggr) C_F^2 T_F\biggr]& 
\end{flalign}

%% file: Coefficients_CO2-VA_wot.tex

\begin{flalign}
%
&c_{2,1}^{V,T} = c_{2,3}^{V,T} = c_{2,1}^{V,L} = c_{2,2}^{V,L} = c_{2,3}^{V,L} = c_{2,4}^{V,L} = c_{2,3}^{A,T} = c_{2,2}^{A,L} = c_{2,3}^{A,L} = 0\\[3ex] 
%
%
%
& c_{2,4}^{V,T} = \left(\frac{\alpha_s}{\pi}\right)^{\!3}\biggl[\biggl(\frac{4}{3}
-10 \zeta_3\biggr) \frac{\dFFF}{N_c}\biggr]&\\[3ex]
& c_{2,1}^{A,T} = \left(\frac{\alpha_s}{\pi}\right)^{\!3}\biggl[\biggl(\frac{L_q}{2}
-\frac{25}{12}\biggr) C_F T_F^2\biggr]&\\[3ex] 
& c_{2,4}^{A,T} = \left(\frac{\alpha_s}{\pi}\right)^{\!2}\biggl[\biggl(2 \zeta_3
+\frac{3 L_q}{2}
+\frac{4}{3}\biggr) C_F T_F\biggr]
+\left(\frac{\alpha_s}{\pi}\right)^{\!3}\biggl[\biggl(\frac{113 \zeta_3}{18}
+\frac{5 \zeta_5}{3}
+\frac{11 L_q^2}{8}
+\biggl(\frac{11 \zeta_3}{3}
+\frac{389}{72}\biggr) L_q
&\notag\\&\quad
+\frac{7273}{864}\biggr) C_A C_F T_F
+\biggl(-\frac{20 \zeta_3}{9}
-\frac{L_q^2}{2}
+\biggl(-\frac{4 \zeta_3}{3}
-\frac{19}{18}\biggr) L_q
-\frac{431}{216}\biggr) C_F n_f T_F^2
+\biggl(4 \zeta_3
-10 \zeta_5
+\frac{3 L_q}{2}
&\notag\\&\quad
+\frac{123}{32}\biggr) C_F^2 T_F\biggr]&\\[3ex] 
& c_{2,1}^{A,L} = \left(\frac{\alpha_s}{\pi}\right)^{\!3}\biggl[(-3 L_q
-1) C_F T_F^2\biggr]&\\[3ex] 
& c_{2,4}^{A,L} = \left(\frac{\alpha_s}{\pi}\right)^{\!2}\biggl[(-3 L_q
-9) C_F T_F\biggr]
+\left(\frac{\alpha_s}{\pi}\right)^{\!3}\biggl[\biggl(\frac{9 \zeta_3}{2}
-\frac{11 L_q^2}{4}
-\frac{269 L_q}{12}
-\frac{6539}{144}\biggr) C_A C_F T_F
+\biggl(L_q^2
+\frac{19 L_q}{3}
&\notag\\&\quad
+\frac{373}{36}\biggr) C_F n_f T_F^2
+\biggl(9 \zeta_3
+\frac{9 L_q}{4}
+\frac{69}{16}\biggr) C_F^2 T_F\biggr]&\\[3ex] 
& c_{2,2}^{V,T} = c_{2,2}^{A,T} = \left(\frac{\alpha_s}{\pi}\right)^{\!2}\biggl[\biggl(2 \zeta_3
+\frac{L_q}{2}
-\frac{3}{2}\biggr) C_F T_F\biggr]
+\left(\frac{\alpha_s}{\pi}\right)^{\!3}\biggl[\biggl(\frac{16 \zeta_3}{9}
+\frac{5 \zeta_5}{3}
+\frac{11 L_q^2}{24}
+\biggl(\frac{11 \zeta_3}{3}
-\frac{83}{72}\biggr) L_q
&\notag\\&\quad
+\frac{97}{216}\biggr) C_A C_F T_F
+\biggl(-\frac{2 \zeta_3}{9}
-\frac{L_q^2}{6}
+\biggl(\frac{13}{18}
-\frac{4 \zeta_3}{3}\biggr) L_q
-\frac{323}{216}\biggr) C_F n_f T_F^2
+\biggl(7 \zeta_3
-10 \zeta_5
+\frac{157}{96}\biggr) C_F^2 T_F\biggr]&\\[3ex] 
& c_{2,5}^{V,T} = -c_{2,5}^{A,T} = \biggl[1\biggr]
+\frac{\alpha_s}{\pi}\biggl[C_F\biggr]
+\left(\frac{\alpha_s}{\pi}\right)^{\!2}\biggl[\biggl(\frac{11 L_q}{12}
+\frac{83}{72}\biggr) C_A C_F
+\biggl(-\frac{L_q}{3}
-\frac{7}{18}\biggr) C_F n_f T_F
+\frac{27}{8} C_F^2\biggr]
&\notag\\&\quad
+\left(\frac{\alpha_s}{\pi}\right)^{\!3}\biggl[\biggl(-\frac{\zeta_3}{12}
-\frac{11 L_q^2}{18}
-\frac{205 L_q}{108}
-\frac{2039}{648}\biggr) C_A C_F n_f T_F
+\biggl(-\frac{23 \zeta_3}{6}
+\frac{121 L_q^2}{144}
+\frac{1219 L_q}{432}
+\frac{7003}{1296}\biggr) C_A^2 C_F
&\notag\\&\quad
+\biggl(\frac{16 \zeta_3}{3}
+\frac{99 L_q}{16}
+\frac{1937}{144}\biggr) C_A C_F^2
+\biggl(\frac{L_q^2}{9}
+\frac{7 L_q}{27}
+\frac{47}{162}\biggr) C_F n_f^2 T_F^2
+\biggl(\frac{4 \zeta_3}{3}
-\frac{5 L_q}{2}
-\frac{643}{144}\biggr) C_F^2 n_f T_F
&\notag\\&\quad
+\biggl(\frac{77}{32}
-\frac{9 \zeta_3}{2}\biggr) C_F^3\biggr]&\\[3ex] 
& c_{2,6}^{V,T} = c_{2,6}^{A,T} = \frac{\alpha_s}{\pi}\biggl[-\frac{3}{4} C_F\biggr]
+\left(\frac{\alpha_s}{\pi}\right)^{\!2}\biggl[\biggl(-\frac{11 L_q}{16}
-\frac{3}{2}\biggr) C_A C_F
+\biggl(\frac{L_q}{4}
+\frac{1}{4}\biggr) C_F n_f T_F
+\frac{21}{32} C_F^2\biggr]
&\notag\\&\quad
+\left(\frac{\alpha_s}{\pi}\right)^{\!3}\biggl[\biggl(\frac{11 \zeta_3}{12}
-\frac{5 \zeta_5}{6}
+\frac{11 L_q^2}{24}
+\frac{85 L_q}{48}
+\frac{2845}{864}\biggr) C_A C_F n_f T_F
+\biggl(-\frac{4 \zeta_3}{3}
+\frac{55 \zeta_5}{24}
-\frac{121 L_q^2}{192}
-\frac{105 L_q}{32}
&\notag\\&\quad
-\frac{1735}{216}\biggr) C_A^2 C_F
+\biggl(\frac{47 \zeta_3}{12}
+\frac{77 L_q}{64}
+\frac{2309}{288}\biggr) C_A C_F^2
+\biggl(-\frac{L_q^2}{12}
-\frac{L_q}{6}
-\frac{67}{216}\biggr) C_F n_f^2 T_F^2
+\biggl(-\frac{13 \zeta_3}{12}
-\frac{L_q}{4}
&\notag\\&\quad
-\frac{173}{576}\biggr) C_F^2 n_f T_F
+\biggl(-\frac{5 \zeta_3}{2}
-\frac{2185}{384}\biggr) C_F^3\biggr]&\\[3ex] 
& c_{2,5}^{V,L} = -c_{2,6}^{V,L} = -c_{2,5}^{A,L} = -c_{2,6}^{A,L} = \biggl[-1\biggr]& 
\end{flalign}

%% file: Coefficients_CO2-SP_wot.tex

\begin{flalign}
%
&c_{2,1}^{S} = c_{2,4}^{S} = c_{2,1}^{P} = c_{2,4}^{P} = 0\\[3ex] 
%
%
%
& c_{2,3}^{S} = \left(\frac{\alpha_s}{\pi}\right)^{\!2}\biggl[\biggl(\frac{9}{2}
-9 \zeta_3\biggr) C_F T_F\biggr]
+\left(\frac{\alpha_s}{\pi}\right)^{\!3}\biggl[\biggl(-\frac{407 \zeta_3}{8}
-\frac{165 \zeta_5}{8}
+\biggl(\frac{33}{4}
-\frac{33 \zeta_3}{2}\biggr) L_q
+\frac{781}{24}\biggr) C_A C_F T_F
&\notag\\&\quad
+\biggl(14 \zeta_3
+\biggl(6 \zeta_3
-3\biggr) L_q
-\frac{25}{3}\biggr) C_F n_f T_F^2
+\biggl(-\frac{87 \zeta_3}{2}
+45 \zeta_5
+\biggl(\frac{27}{4}
-\frac{27 \zeta_3}{2}\biggr) L_q
+\frac{47}{4}\biggr) C_F^2 T_F\biggr]&\\[3ex] 
& c_{2,3}^{P} = \left(\frac{\alpha_s}{\pi}\right)^{\!2}\biggl[\biggl(9 \zeta_3
+\frac{3}{2}\biggr) C_F T_F\biggr]
+\left(\frac{\alpha_s}{\pi}\right)^{\!3}\biggl[\biggl(\frac{343 \zeta_3}{8}
+\frac{165 \zeta_5}{8}
+\biggl(\frac{33 \zeta_3}{2}
+\frac{11}{4}\biggr) L_q
+\frac{185}{24}\biggr) C_A C_F T_F
&\notag\\&\quad
+\biggl(-10 \zeta_3
+\biggl(-6 \zeta_3
-1\biggr) L_q
-\frac{5}{3}\biggr) C_F n_f T_F^2
+\biggl(\frac{75 \zeta_3}{2}
-45 \zeta_5
+\biggl(\frac{27 \zeta_3}{2}
+\frac{9}{4}\biggr) L_q
-\frac{3}{4}\biggr) C_F^2 T_F\biggr]&\\[3ex] 
& c_{2,2}^{S} = c_{2,2}^{P} = \left(\frac{\alpha_s}{\pi}\right)^{\!2}\biggl[\biggl(\frac{3 L_q}{4}
-\frac{5}{4}\biggr) C_F T_F\biggr]
+\left(\frac{\alpha_s}{\pi}\right)^{\!3}\biggl[\biggl(-\frac{43 \zeta_3}{8}
+\frac{11 L_q^2}{16}
+\frac{41 L_q}{48}
-\frac{29}{16}\biggr) C_A C_F T_F
+\biggl(\zeta_3
-\frac{L_q^2}{4}
&\notag\\&\quad
+\frac{5 L_q}{12}
-\frac{15}{16}\biggr) C_F n_f T_F^2
+\biggl(\frac{9 \zeta_3}{4}
+\frac{9 L_q^2}{8}
-\frac{9 L_q}{8}
+\frac{17}{64}\biggr) C_F^2 T_F\biggr]&\\[3ex] 
& c_{2,5}^{S} = -c_{2,5}^{P} = \biggl[1\biggr]
+\frac{\alpha_s}{\pi}\biggl[\biggl(\frac{3 L_q}{2}
+\frac{7}{2}\biggr) C_F\biggr]
+\left(\frac{\alpha_s}{\pi}\right)^{\!2}\biggl[\biggl(\frac{9 \zeta_3}{8}
+\frac{11 L_q^2}{16}
+\frac{251 L_q}{48}
+\frac{625}{64}\biggr) C_A C_F
+\biggl(-\frac{L_q^2}{4}
-\frac{19 L_q}{12}
&\notag\\&\quad
-\frac{41}{16}\biggr) C_F n_f T_F
+\biggl(-\frac{27 \zeta_3}{4}
+\frac{9 L_q^2}{8}
+\frac{87 L_q}{16}
+\frac{481}{64}\biggr) C_F^2\biggr]
+\left(\frac{\alpha_s}{\pi}\right)^{\!3}\biggl[\biggl(-\frac{197 \zeta_3}{72}
-\frac{11 L_q^3}{36}
-\frac{505 L_q^2}{144}
+\biggl(-\frac{9 \zeta_3}{4}
&\notag\\&\quad
-\frac{2875}{216}\biggr) L_q
-\frac{\pi ^4}{120}
-\frac{151907}{7776}\biggr) C_A C_F n_f T_F
+\biggl(-\frac{163 \zeta_3}{8}
-\frac{615 \zeta_5}{16}
+\frac{33 L_q^3}{32}
+\frac{667 L_q^2}{64}
+\biggl(\frac{4415}{128}
-\frac{171 \zeta_3}{16}\biggr) L_q
&\notag\\&\quad
+\frac{24251}{768}\biggr) C_A C_F^2
+\biggl(-\frac{41 \zeta_3}{144}
+\frac{75 \zeta_5}{8}
+\frac{121 L_q^3}{288}
+\frac{3067 L_q^2}{576}
+\biggl(\frac{33 \zeta_3}{16}
+\frac{1279}{54}\biggr) L_q
+\frac{2531137}{62208}\biggr) C_A^2 C_F
+\biggl(\frac{\zeta_3}{9}
+\frac{L_q^3}{18}
&\notag\\&\quad
+\frac{19 L_q^2}{36}
+\frac{167 L_q}{108}
+\frac{6853}{3888}\biggr) C_F n_f^2 T_F^2
+\biggl(\frac{21 \zeta_3}{2}
-\frac{3 L_q^3}{8}
-\frac{7 L_q^2}{2}
+\biggl(6 \zeta_3
-\frac{101}{8}\biggr) L_q
+\frac{\pi ^4}{120}
-\frac{519}{32}\biggr) C_F^2 n_f T_F
&\notag\\&\quad
+\biggl(-\frac{229 \zeta_3}{8}
+\frac{315 \zeta_5}{8}
+\frac{9 L_q^3}{16}
+\frac{135 L_q^2}{32}
+\biggl(\frac{1785}{128}
-\frac{81 \zeta_3}{8}\biggr) L_q
+\frac{2327}{96}\biggr) C_F^3\biggr]&\\[3ex] 
& c_{2,6}^{S} = c_{2,6}^{P} = \biggl[\frac{1}{2}\biggr]
+\frac{\alpha_s}{\pi}\biggl[\biggl(\frac{3 L_q}{4}
+\frac{11}{8}\biggr) C_F\biggr]
+\left(\frac{\alpha_s}{\pi}\right)^{\!2}\biggl[\biggl(-\frac{9 \zeta_3}{16}
+\frac{11 L_q^2}{32}
+\frac{109 L_q}{48}
+\frac{1307}{384}\biggr) C_A C_F
+\biggl(-\frac{L_q^2}{8}
-\frac{2 L_q}{3}
&\notag\\&\quad
-\frac{79}{96}\biggr) C_F n_f T_F
+\biggl(-\frac{9 \zeta_3}{8}
+\frac{9 L_q^2}{16}
+\frac{69 L_q}{32}
+\frac{379}{128}\biggr) C_F^2\biggr]
+\left(\frac{\alpha_s}{\pi}\right)^{\!3}\biggl[\biggl(\frac{91 \zeta_3}{144}
-\frac{11 L_q^3}{72}
-\frac{439 L_q^2}{288}
+\biggl(-\frac{3 \zeta_3}{8}
&\notag\\&\quad
-\frac{4037}{864}\biggr) L_q
-\frac{\pi ^4}{240}
-\frac{51523}{7776}\biggr) C_A C_F n_f T_F
+\biggl(-\frac{137 \zeta_3}{64}
-\frac{75 \zeta_5}{32}
+\frac{33 L_q^3}{64}
+\frac{71 L_q^2}{16}
+\biggl(\frac{3279}{256}
-\frac{93 \zeta_3}{32}\biggr) L_q
&\notag\\&\quad
+\frac{2007}{256}\biggr) C_A C_F^2
+\biggl(-\frac{1787 \zeta_3}{288}
+\frac{45 \zeta_5}{32}
+\frac{121 L_q^3}{576}
+\frac{169 L_q^2}{72}
+\biggl(\frac{15319}{1728}
-\frac{33 \zeta_3}{32}\biggr) L_q
+\frac{1783849}{124416}\biggr) C_A^2 C_F
+\biggl(\frac{\zeta_3}{18}
&\notag\\&\quad
+\frac{L_q^3}{36}
+\frac{2 L_q^2}{9}
+\frac{101 L_q}{216}
+\frac{4279}{7776}\biggr) C_F n_f^2 T_F^2
+\biggl(\frac{13 \zeta_3}{8}
-\frac{3 L_q^3}{16}
-\frac{47 L_q^2}{32}
+\biggl(\frac{3 \zeta_3}{2}
-\frac{155}{32}\biggr) L_q
+\frac{\pi ^4}{240}
-\frac{229}{48}\biggr) C_F^2 n_f T_F
&\notag\\&\quad
+\biggl(-\frac{185 \zeta_3}{32}
+\frac{45 \zeta_5}{16}
+\frac{9 L_q^3}{32}
+\frac{27 L_q^2}{16}
+\biggl(\frac{1461}{256}
-\frac{27 \zeta_3}{16}\biggr) L_q
+\frac{15145}{1536}\biggr) C_F^3\biggr]& 
\end{flalign}

%% file: Coefficients_CO6-VA_wot.tex

\begin{flalign}
%
&c_{6,1}^{V,T} = c_{6,3}^{V,T} = c_{6,4}^{V,T} = c_{6,1}^{V,L} = c_{6,2}^{V,L} = c_{6,3}^{V,L} = c_{6,4}^{V,L} = c_{6,4}^{A,T} = c_{6,2}^{A,L} = c_{6,4}^{A,L} = 0\\[3ex] 
%
%
%
& c_{6,1}^{A,T} = \left(\frac{\alpha_s}{\pi}\right)^{\!2}\biggl[\biggl(-\frac{3 \zeta_3}{4}
-\frac{3 L_q^2}{8}
+\biggl(-\frac{\zeta_3}{2}
-\frac{7}{48}\biggr) L_q
+\frac{3}{8}\biggr) C_F T_F\biggr]&\\[3ex] 
& c_{6,3}^{A,T} = \left(\frac{\alpha_s}{\pi}\right)^{\!2}\biggl[\biggl(\frac{20 \zeta_5}{3}
-\frac{5 \zeta_3}{3}\biggr) C_F T_F\biggr]&\\[3ex] 
& c_{6,1}^{A,L} = \left(\frac{\alpha_s}{\pi}\right)^{\!2}\biggl[\biggl(\frac{3 L_q^2}{4}
+\frac{9 L_q}{4}
+\frac{3}{4}\biggr) C_F T_F\biggr]&\\[3ex] 
& c_{6,3}^{A,L} = \left(\frac{\alpha_s}{\pi}\right)^{\!2}\biggl[\biggl(-4 \zeta_3
-5 \zeta_5\biggr) C_F T_F\biggr]&\\[3ex] 
& c_{6,2}^{V,T} = c_{6,2}^{A,T} = \left(\frac{\alpha_s}{\pi}\right)^{\!2}\biggl[\biggl(\frac{\zeta_3}{2}
-\frac{L_q^2}{16}
+\biggl(\frac{5}{12}
-\frac{\zeta_3}{2}\biggr) L_q
-\frac{5}{6}\biggr) C_F T_F\biggr]&\\[3ex] 
& c_{6,5}^{V,T} = c_{6,5}^{A,T} = \biggl[-\frac{1}{8}\biggr]
+\frac{\alpha_s}{\pi}\biggl[\biggl(-\frac{3 L_q}{16}
-\frac{3}{16}\biggr) C_F\biggr]
+\left(\frac{\alpha_s}{\pi}\right)^{\!2}\biggl[\biggl(\frac{61 \zeta_3}{48}
-\frac{5 \zeta_5}{4}
-\frac{29 L_q}{96}
-\frac{227}{576}\biggr) C_A C_F
+\biggl(-\frac{\zeta_3}{24}
+\frac{5 L_q}{48}
&\notag\\&\quad
+\frac{23}{288}\biggr) C_F n_f T_F
+\biggl(\frac{25 \zeta_3}{24}
-\frac{5 \zeta_5}{3}
-\frac{9 L_q^2}{32}
-\frac{75 L_q}{128}
-\frac{95}{768}\biggr) C_F^2\biggr]&\\[3ex] 
& c_{6,6}^{V,T} = c_{6,6}^{A,T} = \left(\frac{\alpha_s}{\pi}\right)^{\!2}\biggl[\biggl(\frac{25}{24}
-\frac{\zeta_3}{2}\biggr) C_F T_F\biggr]&\\[3ex] 
& c_{6,7}^{V,T} = c_{6,7}^{A,T} = \frac{\alpha_s}{\pi}\biggl[\biggl(\frac{19}{24}
-\frac{\zeta_3}{2}\biggr) C_F\biggr]
+\left(\frac{\alpha_s}{\pi}\right)^{\!2}\biggl[\biggl(-\frac{209 \zeta_3}{144}
+\frac{5 \zeta_5}{24}
+\biggl(\frac{209}{288}
-\frac{11 \zeta_3}{24}\biggr) L_q
+\frac{2537}{1728}\biggr) C_A C_F
&\notag\\&\quad
+\biggl(\frac{\zeta_3}{36}
+\biggl(\frac{\zeta_3}{6}
-\frac{19}{72}\biggr) L_q
-\frac{181}{432}\biggr) C_F n_f T_F
+\biggl(-\frac{23 \zeta_3}{8}
+\frac{15 \zeta_5}{4}
+\biggl(\frac{19}{8}
-\frac{3 \zeta_3}{2}\biggr) L_q
+\frac{337}{192}\biggr) C_F^2\biggr]&\\[3ex] 
& c_{6,8}^{V,T} = -c_{6,8}^{A,T} = \biggl[-\frac{L_q}{4}\biggr]
+\frac{\alpha_s}{\pi}\biggl[\biggl(\frac{3 \zeta_3}{4}
-\frac{3 L_q^2}{8}
-\frac{L_q}{2}
-\frac{7}{8}\biggr) C_F\biggr]
+\left(\frac{\alpha_s}{\pi}\right)^{\!2}\biggl[\biggl(\frac{37 \zeta_3}{16}
+\frac{5 \zeta_5}{24}
-\frac{11 L_q^3}{96}
-\frac{163 L_q^2}{192}
+\biggl(\frac{17 \zeta_3}{16}
&\notag\\&\quad
-\frac{2237}{1152}\biggr) L_q
+\frac{\pi ^4}{480}
-\frac{6761}{2304}\biggr) C_A C_F
+\biggl(-\frac{5 \zeta_3}{12}
+\frac{L_q^3}{24}
+\frac{11 L_q^2}{48}
+\biggl(\frac{157}{288}
-\frac{\zeta_3}{4}\biggr) L_q
+\frac{337}{576}\biggr) C_F n_f T_F
+\biggl(\frac{37 \zeta_3}{48}
-\frac{5 \zeta_5}{6}
&\notag\\&\quad
-\frac{3 L_q^3}{8}
-\frac{51 L_q^2}{64}
+\biggl(\frac{3 \zeta_3}{2}
-\frac{345}{128}\biggr) L_q
-\frac{\pi ^4}{240}
+\frac{329}{768}\biggr) C_F^2\biggr]&\\[3ex] 
& c_{6,9}^{V,T} = -c_{6,9}^{A,T} = \left(\frac{\alpha_s}{\pi}\right)^{\!2}\biggl[\biggl(\zeta_3
+\frac{3 L_q}{2}
-\frac{2}{3}\biggr) C_F T_F\biggr]&\\[3ex] 
& c_{6,5}^{V,L} = c_{6,5}^{A,L} = \biggl[-\frac{L_q}{4}
-\frac{1}{4}\biggr]
+\frac{\alpha_s}{\pi}\biggl[\biggl(\frac{3 \zeta_3}{4}
-\frac{3 L_q^2}{8}
-L_q
-\frac{25}{16}\biggr) C_F\biggr]
+\left(\frac{\alpha_s}{\pi}\right)^{\!2}\biggl[\biggl(\frac{157 \zeta_3}{96}
+\frac{5 \zeta_5}{16}
-\frac{11 L_q^3}{96}
-\frac{185 L_q^2}{192}
&\notag\\&\quad
+\biggl(\frac{17 \zeta_3}{16}
-\frac{1265}{384}\biggr) L_q
+\frac{\pi ^4}{480}
-\frac{9469}{2304}\biggr) C_A C_F
+\biggl(-\frac{5 \zeta_3}{24}
+\frac{L_q^3}{24}
+\frac{13 L_q^2}{48}
+\biggl(\frac{85}{96}
-\frac{\zeta_3}{4}\biggr) L_q
+\frac{605}{576}\biggr) C_F n_f T_F
&\notag\\&\quad
+\biggl(\frac{137 \zeta_3}{32}
-\frac{25 \zeta_5}{8}
-\frac{3 L_q^3}{8}
-\frac{99 L_q^2}{64}
+\biggl(\frac{3 \zeta_3}{2}
-\frac{481}{128}\biggr) L_q
-\frac{\pi ^4}{240}
-\frac{2179}{768}\biggr) C_F^2\biggr]&\\[3ex] 
& c_{6,6}^{V,L} = -\frac{1}{2} c_{6,9}^{V,L} = c_{6,6}^{A,L} = \frac{c_{6,9}^{A,L}}{2} = \left(\frac{\alpha_s}{\pi}\right)^{\!2}\biggl[\biggl(\frac{3 L_q}{4}
+\frac{1}{3}\biggr) C_F T_F\biggr]&\\[3ex] 
& c_{6,7}^{V,L} = c_{6,7}^{A,L} = \biggl[\frac{1}{2}\biggr]
+\frac{\alpha_s}{\pi}\biggl[\biggl(\frac{3 L_q}{2}
+\frac{19}{8}\biggr) C_F\biggr]
+\left(\frac{\alpha_s}{\pi}\right)^{\!2}\biggl[\biggl(-\frac{23 \zeta_3}{16}
+\frac{11 L_q^2}{16}
+\frac{403 L_q}{96}
+\frac{787}{96}\biggr) C_A C_F
+\biggl(\frac{\zeta_3}{4}
-\frac{L_q^2}{4}
&\notag\\&\quad
-\frac{29 L_q}{24}
-\frac{53}{24}\biggr) C_F n_f T_F
+\biggl(-\frac{29 \zeta_3}{16}
-\frac{5 \zeta_5}{2}
+\frac{9 L_q^2}{4}
+\frac{117 L_q}{16}
+\frac{323}{32}\biggr) C_F^2\biggr]&\\[3ex] 
& c_{6,8}^{V,L} = -c_{6,8}^{A,L} = \biggl[\frac{L_q}{4}\biggr]
+\frac{\alpha_s}{\pi}\biggl[\biggl(-\frac{3 \zeta_3}{4}
+\frac{3 L_q^2}{8}
+\frac{L_q}{4}
+\frac{3}{8}\biggr) C_F\biggr]
+\left(\frac{\alpha_s}{\pi}\right)^{\!2}\biggl[\biggl(-\frac{11 \zeta_3}{12}
-\frac{5 \zeta_5}{16}
+\frac{11 L_q^3}{96}
+\frac{119 L_q^2}{192}
+\biggl(\frac{153}{128}
&\notag\\&\quad
-\frac{17 \zeta_3}{16}\biggr) L_q
-\frac{\pi ^4}{480}
+\frac{25}{2304}\biggr) C_A C_F
+\biggl(\frac{\zeta_3}{12}
-\frac{L_q^3}{24}
-\frac{7 L_q^2}{48}
+\biggl(\frac{\zeta_3}{4}
-\frac{9}{32}\biggr) L_q
+\frac{31}{576}\biggr) C_F n_f T_F
+\biggl(-\frac{27 \zeta_3}{8}
+\frac{35 \zeta_5}{8}
&\notag\\&\quad
+\frac{3 L_q^3}{8}
+\frac{27 L_q^2}{64}
+\biggl(\frac{13}{128}
-\frac{3 \zeta_3}{2}\biggr) L_q
+\frac{\pi ^4}{240}
-\frac{1697}{768}\biggr) C_F^2\biggr]& 
&\notag\\&
= - c_{6,5}^{V,L} - \frac{1}{2} c_{6,7}^{V,L}
\end{flalign}

%% file: Coefficients_CO6-SP_wot.tex

\begin{flalign}
%
&c_{6,1}^{S} = c_{6,3}^{S} = c_{6,1}^{P} = c_{6,3}^{P} = 0\\[3ex] 
%
%
%
& c_{6,4}^{S} = \left(\frac{\alpha_s}{\pi}\right)^{\!2}\biggl[\biggl(2 \zeta_3
-5 \zeta_5
+\biggl(\frac{9 \zeta_3}{4}
-\frac{9}{8}\biggr) L_q
+\frac{3}{2}\biggr) C_F T_F\biggr]&\\[3ex] 
& c_{6,4}^{P} = \left(\frac{\alpha_s}{\pi}\right)^{\!2}\biggl[\biggl(-3 \zeta_3
+\biggl(-\frac{9 \zeta_3}{4}
-\frac{3}{8}\biggr) L_q
+1\biggr) C_F T_F\biggr]&\\[3ex] 
& c_{6,2}^{S} = c_{6,2}^{P} = \left(\frac{\alpha_s}{\pi}\right)^{\!2}\biggl[\biggl(\frac{3 \zeta_3}{8}
-\frac{3 L_q^2}{32}
+\frac{3 L_q}{8}
-\frac{15}{32}\biggr) C_F T_F\biggr]&\\[3ex] 
& c_{6,5}^{S} = c_{6,5}^{P} = \biggl[\frac{1}{16}
-\frac{L_q}{8}\biggr]
+\frac{\alpha_s}{\pi}\biggl[\biggl(\frac{3 \zeta_3}{8}
-\frac{3 L_q^2}{8}
-\frac{3 L_q}{16}
-\frac{9}{32}\biggr) C_F\biggr]
+\left(\frac{\alpha_s}{\pi}\right)^{\!2}\biggl[\biggl(\frac{107 \zeta_3}{384}
+\frac{15 \zeta_5}{32}
-\frac{55 L_q^3}{384}
-\frac{575 L_q^2}{768}
&\notag\\&\quad
+\biggl(\frac{43 \zeta_3}{64}
-\frac{1775}{1536}\biggr) L_q
+\frac{\pi ^4}{960}
-\frac{3977}{9216}\biggr) C_A C_F
+\biggl(\frac{5 \zeta_3}{24}
+\frac{5 L_q^3}{96}
+\frac{37 L_q^2}{192}
+\biggl(\frac{115}{384}
-\frac{\zeta_3}{8}\biggr) L_q
+\frac{205}{2304}\biggr) C_F n_f T_F
&\notag\\&\quad
+\biggl(\frac{9 \zeta_3}{32}
-\frac{5 \zeta_5}{16}
-\frac{39 L_q^3}{64}
-\frac{105 L_q^2}{128}
+\biggl(\frac{51 \zeta_3}{32}
-\frac{923}{512}\biggr) L_q
-\frac{\pi ^4}{480}
-\frac{353}{3072}\biggr) C_F^2\biggr]&\\[3ex] 
& c_{6,6}^{S} = c_{6,6}^{P} = \left(\frac{\alpha_s}{\pi}\right)^{\!2}\biggl[\biggl(-\frac{3 \zeta_3}{4}
+\frac{3 L_q}{8}
-\frac{43}{48}\biggr) C_F T_F\biggr]&\\[3ex] 
& c_{6,7}^{S} = c_{6,7}^{P} = \biggl[\frac{1}{4}\biggr]
+\frac{\alpha_s}{\pi}\biggl[\biggl(\frac{3 \zeta_3}{4}
+\frac{9 L_q}{8}
+\frac{1}{2}\biggr) C_F\biggr]
+\left(\frac{\alpha_s}{\pi}\right)^{\!2}\biggl[\biggl(-\frac{41 \zeta_3}{48}
+\frac{15 \zeta_5}{32}
+\frac{33 L_q^2}{64}
+\biggl(\frac{11 \zeta_3}{16}
+\frac{379}{192}\biggr) L_q
&\notag\\&\quad
+\frac{7223}{2304}\biggr) C_A C_F
+\biggl(\frac{\zeta_3}{3}
-\frac{3 L_q^2}{16}
+\biggl(-\frac{\zeta_3}{4}
-\frac{23}{48}\biggr) L_q
-\frac{463}{576}\biggr) C_F n_f T_F
+\biggl(\frac{19 \zeta_3}{4}
-\frac{55 \zeta_5}{16}
+\frac{81 L_q^2}{32}
&\notag\\&\quad
+\biggl(\frac{27 \zeta_3}{8}
+\frac{153}{64}\biggr) L_q
+\frac{683}{256}\biggr) C_F^2\biggr]&\\[3ex] 
& c_{6,8}^{S} = -c_{6,8}^{P} = \biggl[-\frac{L_q}{4}\biggr]
+\frac{\alpha_s}{\pi}\biggl[\biggl(\frac{3 \zeta_3}{4}
-\frac{3 L_q^2}{4}
-\frac{9 L_q}{8}
-\frac{3}{4}\biggr) C_F\biggr]
+\left(\frac{\alpha_s}{\pi}\right)^{\!2}\biggl[\biggl(\frac{31 \zeta_3}{96}
+\frac{25 \zeta_5}{32}
-\frac{55 L_q^3}{192}
-\frac{185 L_q^2}{96}
+\biggl(\frac{25 \zeta_3}{32}
&\notag\\&\quad
-\frac{1019}{256}\biggr) L_q
+\frac{\pi ^4}{480}
-\frac{4357}{2304}\biggr) C_A C_F
+\biggl(\frac{\zeta_3}{6}
+\frac{5 L_q^3}{48}
+\frac{13 L_q^2}{24}
+\biggl(\frac{67}{64}
-\frac{\zeta_3}{4}\biggr) L_q
+\frac{245}{576}\biggr) C_F n_f T_F
+\biggl(6 \zeta_3
-\frac{25 \zeta_5}{16}
&\notag\\&\quad
-\frac{39 L_q^3}{32}
-\frac{111 L_q^2}{32}
+\biggl(\frac{69 \zeta_3}{16}
-\frac{1307}{256}\biggr) L_q
-\frac{\pi ^4}{240}
-\frac{1483}{768}\biggr) C_F^2\biggr]&\\[3ex] 
& c_{6,9}^{S} = -c_{6,9}^{P} = \left(\frac{\alpha_s}{\pi}\right)^{\!2}\biggl[\biggl(-\frac{3 \zeta_3}{2}
+\frac{3 L_q}{2}
-\frac{1}{3}\biggr) C_F T_F\biggr]& 
\end{flalign}